\documentclass[lettersize,journal]{IEEEtran}
\IEEEoverridecommandlockouts
\usepackage{cite}
\usepackage{amsmath,amssymb,amsfonts}
\usepackage{graphicx}
\usepackage{textcomp}
\usepackage{xcolor}
\usepackage{algorithm}
\usepackage{algpseudocode}

\usepackage{booktabs,array}
\usepackage{subfigure}
\usepackage{bm}
\usepackage{comment}
\usepackage{color}
\usepackage{stfloats}
\usepackage{makecell}
\usepackage{diagbox}
\allowdisplaybreaks[4]
\begin{document}
	
	\title{Joint Beamforming and Antenna Position Optimization for IRS-Aided Multi-User Movable Antenna Systems\\
	}
	
	\author{\IEEEauthorblockN{Yue Geng, \IEEEmembership{Graduate Student Member, IEEE}, Tee Hiang Cheng, Kai Zhong, \\ Kah Chan Teh, \IEEEmembership{Senior Member, IEEE}, and Qingqing Wu, \IEEEmembership{Senior Member, IEEE}}
		\thanks{The work of Yue Geng, Tee Hiang Cheng and Kah Chan Teh was supported by School of Electrical and Electronic Engineering, Nanyang Technological University. The work of Kai Zhong was supported in part by NSFC 62501112, in part by the China Postdoctoral Science Foundation under Grant 2025M773511, and in part by the Municipal Government of Quzhou under Grant 2025K008. The work of Qingqing Wu was supported by NSFC 62371289 and NSFC 62331022.  \emph{Corresponding authors: Kah Chan Teh}.
			
			Yue Geng, Tee Hiang Cheng, Kah Chan Teh are with the School of Electrical and Electronic Engineering, Nanyang Technological University, Singapore (Email: yue014@e.ntu.edu.sg; ethcheng@ntu.edu.sg; ekcteh@ntu.edu.sg).
			
			Kai~Zhong is with the School of Information and Communication Engineering, University of Electronic Science and Technology of China, Chengdu, 611731, China (Email: 201921011206@std.uestc.edu.cn).
			
			Qingqing Wu is with the Department of Electronic Engineering, Shanghai Jiao Tong University, 200240, China (e-mail: qingqingwu@sjtu.edu.cn).
		}

	}
	
	\maketitle
	
	\begin{abstract}
		Intelligent reflecting surface (IRS) and movable antenna (MA) technologies have been proposed to enhance wireless communications by creating favorable channel conditions. This paper investigates the joint beamforming and antenna position optimization for MA-enabled IRS (MA-IRS)-aided multi-user multiple-input single-output (MU-MISO) communication systems, where the MA-IRS is deployed to aid the communication between the MA-enabled base station (BS) and user equipment (UE). In contrast to conventional fixed position antenna (FPA)-enabled IRS (FPA-IRS), the positions of the reflecting elements of the MA-IRS can be controlled to enhances the wireless channel. To verify the system's effectiveness and optimize its performance, we formulate a sum-rate maximization problem with a minimum rate threshold constraint for the MU-MISO communication. To tackle the non-convex problem, a product Riemannian manifold optimization (PRMO) method is proposed for the joint optimization of the beamforming and MA positions. Specifically, a product Riemannian manifold space (PRMS) is constructed and the corresponding Riemannian gradient is derived for updating the variables, and the Riemannian exact penalty (REP) method and a Riemannian Broyden–Fletcher–Goldfarb–Shanno (RBFGS) algorithm is exploited to obtain a feasible solution over the PRMS. Simulation results demonstrate that compared with the conventional FPA-IRS-aided communications, the reflecting elements of the MA-IRS can move to the positions with higher channel gain, thus enhancing the system performance. Furthermore, it is shown that optimizing the positions of the reflecting elements brings higher performance gain than controlling the phase shifts of the IRS, and integrating MA with IRS leads to higher performance gains compared to integrating MA with BS.
	\end{abstract}
	
	\begin{IEEEkeywords}
		Movable antenna (MA), intelligent reflecting surface (IRS), multi-user communication, product Riemannian manifold optimization, Riemannian exact penalty.
	\end{IEEEkeywords}
	
	\section{Introduction}
	Driven by the demand for higher data rates, improved connectivity, and enhanced transmission reliability, multiple-input multiple-output (MIMO) and massive MIMO have emerged as promising technologies for sixth-generation (6G) and beyond wireless communication systems due to their extensive network coverage, cost-effectiveness, and significant energy efficiency (EE) \cite{6g,mimo1}. By utilizing the abundant degrees of freedom (DoF) and beamforming techniques, MIMO systems can effectively suppress interference, thereby minimizing signal disruption and optimizing overall network performance \cite{mimo2}. To further enhance the performance of wireless systems by improving the wireless channels, intelligent reflecting surface (IRS) has emerged as a promising technology for enhancing wireless channel environments. As a metasurface composed of passive reflecting elements, an IRS can adjust the phase shifts of the reflected signals, thus adding coherence with the signals from different paths and enhancing the signal power received by the users \cite{IRS1,IRS2,IRS3}. Due to its capability of steering the signals toward the desired direction, IRS can address the issues of obstacles and fading in conventional massive MIMO, which occur in certain scenarios when the line-of-sight (LoS) links between the base station (BS) and user equipment (UE) are severely blocked. Additionally, IRS can further enhance signal quality, extend coverage, and reduce power consumption. Exploiting the advantages, IRS has been widely studied and applied for enhancing the performance of wireless communication systems \cite{IRS4,IRS5,IRS6,IRS7}. To achieve higher performance gain, typically from dozens to hundreds of reflecting elements are employed for the conventional IRS. This could complicate precise phase control. Although the complexity can be reduced by employing techniques such as discrete phase control, this may lead to performance degradation. Besides, numerous reflecting elements lead to larger surface region size, as a specific distance must be maintained between each element to avoid the coupling effects. This makes it challenging to deploy a large number of IRS elements in certain application scenarios, such as aerial IRS (AIRS), where the IRS is mounted on aerial platforms like unmanned aerial vehicles (UAVs) \cite{AIRS}. This motivate us to improve IRS by advanced techniques to enhance the performance of IRS-aided systems where the size of IRS is limited.
	
	To further create favorable channel conditions, the concept of movable antennas (MA) has been proposed as a novel technique for future wireless systems \cite{MA1,MA2}. To meet the growing quality of service (QoS) requirements, conventional massive MIMO systems can deploy arrays with several hundred antennas to serve multiple users simultaneously within the same time-frequency resources \cite{mimo3}. However, the large number of antennas in conventional fixed position array (FPA) can lead to higher hardware costs and energy consumption due to the increasing demand for radio frequency (RF) chains \cite{mimo4}. Additionally, the unchangeable geometric configurations can result in array gain loss within the large antenna region of the conventional FPA \cite{MAIRSISAC}. To tackle this issue, the antenna selection (AS) technology was proposed, where an optimal antenna subset can be selected from a pre-defined dense array with favorable channel conditions. However, higher hardware cost is required for developing the massive candidate antennas. To reduce the number of radio-frequency (RF) chains and fully exploit the DoF within a confined region, MA was proposed as a new solution to tackle the fundamental limitations of FPA systems. Specifically, the positions of the MA can be flexibly controlled in response to channel conditions to achieve higher channel power gains \cite{MA1,MA2,MA3}. Within a confined region in the order of several wavelengths, MAs can move continuously using mechanical controllers and drivers to adjust the phase of different channel paths, thereby fully leveraging the spatial DoF to enhance wireless channel gains \cite{MA9,MA10,MA11}. Considering the advantages of MA and IRS in wireless communication systems, recent studies have begun to explore the implementation of IRS in MA-enabled wireless systems to further enhance channel conditions \cite{MAIRSISAC, MAIRS1}. 
	
	Exploiting the advantages of MA, recent research has begun to focus on MA-enabled IRS (MA-IRS) \cite{MAIRS2, MAIRS3}. For the MA-IRS, each reflecting element can be installed on a mechanical slide track, which is driven by step motors. The element is connected to controller via a flexible cable. After receiving the signal from the central processing unit (CPU), the phase shift of the reflecting element can be adjusted. Meanwhile, the motors are controlled by the CPU via control link, and can cooperatively relocate the element to a target position within a given region \cite{MA1}. Compared to conventional FPA-enabled IRS (FPA-IRS), the reflecting elements of MA-IRS can be moved to positions with higher channel gain, resulting in enhanced communication performance with fewer reflecting elements. Although adjusting the position of the MA takes time, for example, the response time of a motor-based MA ranges from milliseconds to seconds \cite{rtime}, in many scenarios such as industrial internet of things (IoT), UEs are often static or move slowly, which makes the time overhead for adjusting MA positions tolerable compared to the longer channel coherence time, allowing the MA to effectively enhance system performance \cite{MA6,MA9,MAIRS2}. Besides, the region size of MA-IRS can be controlled in the order of several wavelengths, making it suitable for scenarios with limited equipment size. Some related works regarding MA, IRS, and MA-IRS are reviewed in the next subsection. 
	
	\subsection{Related Works}
	Compared with the conventional MIMO systems, the optimization of the movable antenna position (MAP) is crucial for enhancing the performance of MA-enabled wireless systems. In \cite{MA3}, the channel modelling of the MA-enabled communication system was presented, proving that MA outperforms FPA in terms of maximum channel capacity with proper MA position optimization. For the MA-enabled uplink multi-user communication system, a MAP optimization method was investigated in \cite{MA6}, where the channel functions with respect to the MAP have been derived, and the zero-forcing (ZF) and minimum mean square error (MMSE) based algorithms were proposed for the MAP optimization of BS and UE. An MA-enabled downlink multi-user multiple-input single-output (MU-MISO) communication system was investigated in \cite{MA10}, where a sparse optimization (SO)-based approach focusing on regularized zero-forcing (RZF) was developed to optimize the antenna positions and the precoding matrix for minimizing the inter-user interference and transmit power. For the sum-rate maximization of such a system, the fractional programming (FP) and ZF-based algorithms were proposed in \cite{MA7}, where the gradient descent (GD) method was utilized for updating the MAP. Besides, the weighted MMSE (WMMSE) and majorization-minimization (MM) based algorithms were proposed in \cite{MA8} for the weighted sum-rate maximization of an MA-enabled MU-MISO communication system. The capacity of an MA-enabled MIMO system was investigated in \cite{MA9}, where an alternating optimization (AO) algorithm was proposed to maximize the capacity, and it was verified that the MA system outperforms the FPA system in terms of channel capacity. 
	
	The integration of IRS and MA in a downlink MUSO communication system was investigated in \cite{MAIRS4}, where an AO method was proposed for the optimization, and it was shown that higher channel gain and system performance were achieved by implementing MA. For an IRS-aided MU-MISO communication system where the BS is equipped with MA, an FP-based AO algorithm was proposed in \cite{MAIRS1} to maximize the system's sum-rate, where the precoding matrix of BS, the phase shifts of the IRS, and the MAP of the BS were optimized via FP, MM, and GD alternatively. An MA-enabled IRS-aided integrated sensing and communication (ISAC) system was studied in \cite{MAIRSISAC}, where the IRS phase shift and MAP optimization were tackled by the sequential rank-one constraint relaxation (SRCR) and successive convex approximation (SCA), respectively. For the above works, the IRS was assumed to be the conventional FPA-IRS, and the phase shifts of the IRS were optimized to enhance the system performance. 
	
	Equipping IRS with MA is a newly emerging area of research. For the MA-IRS-aided wireless system, it was demonstrated in \cite{MAIRS2} that the phase distribution offset problem of the conventional FPA-IRS can be addressed if the position of each IRS element can be flexibly adjusted exploiting the MA technology, and an iterative algorithm was developed for designing the discrete phase shifts of the MA-IRS. Moreover, it has been shown in \cite{MAIRS3} that the MA-IRS outperforms the conventional FPA-IRS in terms of signal-to-noise ratio (SNR) and outage probability performance for a multi-user communication system, and the results indicated that the system achieves performance gains by controlling the positions of the IRS elements under conditions where the number of IRS elements is limited and the IRS size is constrained to a few wavelengths.
	
	\subsection{Motivations and Contributions}
	Although some research has been conducted on MA-IRS-aided wireless systems, existing methods typically assume that only the BS or IRS is equipped with MA, and the MAP and other system configurations are optimized alternatively. Motivated by these works, it is intuitive that the channel can be further improved by equipping both the BS and IRS with MA simultaneously, while existing methods cannot tackle the optimization for such a system well. To explore the theoretical performance of the MA-IRS in improving the communication performance compared with conventional FPA-IRS, we explore the joint beamforming and antenna position optimization for MA-IRS-aided MU-MISO communication systems in this paper, focusing on the sum-rate maximization problem under minimum rate constraints. Specifically, we proposed a product Riemannian manifold optimization (PRMO) method for the joint optimization. For the MA-IRS, we investigate two modes. In the first mode, both the position and phase shift of each element can be jointly controlled. In the second mode, the phase shift remains fixed while only the element’s position is adjustable, thereby avoiding the need for precise phase control of numerous reflecting elements and reducing the complexity. Optimization for both the modes can be tackled by the proposed PRMO. The primary contributions of this work are as follows:
	\begin{itemize}
		\item We investigate the MA-IRS-aided MU-MISO communication systems, which have not been extensively studied in the literature. Compared with conventional FPA-IRS, the MA-IRS enhances the wireless channel by controlling the MAPs. We characterize channel models as functions of the MAPs and formulate the signal-to-interference-plus-noise ratio (SINR) obtained by each UE, which is intractable due to the coupling of precoding matrix and the MAPs of the BS and IRS. 
		\item We formulate the joint beamforming and MAP optimization for the system as a sum-rate maximization problem, where a minimum rate threshold for each UE is ensured. To address this highly non-convex problem, we propose a PRMO method for the simultaneous updating of variables. The constraints of the maximum transmit power, the constant modulus of the IRS, and the MA regions are handled by confining the variables to Riemannian manifold spaces (RMS), which are then combined into a product Riemannian manifold space (PRMS) serving as the solution space. The Riemannian exact penalty (REP) method and a Riemannian Broyden–Fletcher–Goldfarb–Shanno (RBFGS) algorithm are exploited to obtain a feasible solution of the problem.
		\item Finally, sufficient simulation results are presented to validate the effectiveness of the MA-IRS-aided MU-MISO system and the proposed PRMO method. The results demonstrate that compared with FPA-IRS, the proposed MA-IRS is able to improve the wireless channel gain by controlling the MAP of the reflecting elements, thus achieving a higher minimum rate-constrained sum-rate. It is shown that optimizing the position of MA-IRS elements yields greater performance gains compared to optimizing the phase shifts when the number of reflecting element is limited. Moreover, with optimized element positions, the additional performance improvement from phase shift optimization is minimal, indicating that position optimization with fixed phase shifts can be employed to reduce computational overhead. It is also shown that equipping the IRS with MA results in greater performance gains than equipping the BS with MA. 
	\end{itemize}
	
	The rest of the paper is organized as follows. Section II provides the system model and problem formulation. Section III develops the PRMO method for addressing the non-convex sum-rate maximization problem. Simulation results are presented in Section IV to verify the effectiveness of the MA-IRS and proposed algorithms. The conclusion of the paper is given in Section V.
	
	\textit{Notations:}  Scalars, vectors, and matrices are indicated as $a$, $\mathbf{a}$, and $\mathbf{A}$, respectively. $\mathbf{A}^T$, $\mathbf{A}^H$ and $\mathbf{A}^*$ indicate the transpose, conjugate transpose, and conjugate of $\mathbf{A}$, respectively. $\Re(\cdot)$ denotes taking the real part. $\mathbf{I}$ indicates identity matrix. $\odot$ denotes the element-wise multiplication. $\lvert \cdot \rvert$, $\lVert \cdot \rVert_2$, and $\lVert \cdot \rVert$ indicate modulus, 2-norm and Frobenius norm, respectively. $\operatorname{diag}(\cdot)$ is the diagonal of a vector, $\operatorname{vec}(\cdot)$ and $\operatorname{Tr}(\cdot)$ are the vectorization and trace of a matrix, respectively.

	\section{System Model and Problem Formulation}
	\subsection{System Model}
	
	As shown in Fig. 1, we consider an MA-IRS-aided MU-MISO downlink communication system. The BS equipped with $M$ MAs transmits signals to $K$ single-antenna UEs simultaneously. Since the direct BS-UE channel is blocked, an IRS with $N$ movable reflecting elements is deployed to provide cascaded channel links. The MAs of BS are arranged in a linear array, while those of IRS are arranged in a planar array. For the BS, each MA is connected to the CPU via flexible RF chain and deployed on an one-dimensional mechanical slide track. The slide tracks are driven by step motors, which are connected to CPU via control link and enable controllable MA positions \cite{MA1,rtime}. Similarly, the reflecting elements of the IRS are installed on two-dimensional movable slide tracks driven by controllable motor and connected to the IRS controller via control circuit \cite{IRS1}. Denote the sets of UEs as $\mathcal{K}=\{1,\dots,K\}$ and the sets of MAs of BS and IRS as $\mathcal{B}=\{1,\dots,M\}$ and $\mathcal{N}=\{1,\dots,N\}$, respectively. The positions of the $m$-th MA of BS and $n$-th MA of IRS are denoted as coordinates $t_m = x_{B,m}$ and $\mathbf{u}_n = [x_{I,n},y_{I,n}]^T$, respectively. The sets of MAPs of BS and IRS are given as $\mathbf{t}=[t_1,\dots,t_M]^T\in\mathbb{R}^{M}$ and $\mathbf{u}=[\mathbf{u}^T_1,\dots,\mathbf{u}^T_N]^T\in\mathbb{R}^{2N}$, respectively.
	The MAs of the BS and IRS can move freely in the one-dimensional region $\mathcal{C}_B$ and the two-dimensional regions $\mathcal{C}_I$ with edge lengths of $A_B$ and $A_I$, respectively. To avoid the coupling effect between the antennas, the MAPs of BS and IRS should satisfy $\lvert t_m-t_{m'} \rvert\geq \lambda/2$ and $\lVert \mathbf{u}_n-\mathbf{u}_{n'} \rVert\geq \lambda/2$, where $\lambda$ is the signal wavelength. 
	\begin{figure}[t]
		\centering{\includegraphics[width=1.0\columnwidth]{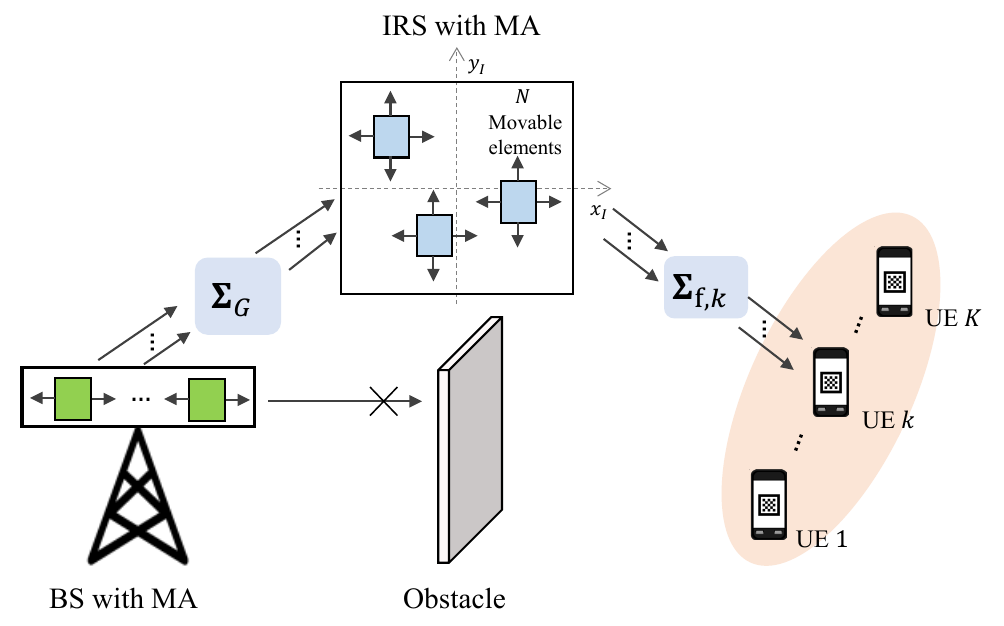}}
		\caption{The MA-IRS-aided MU-MISO communication system.}
	\end{figure}
	
	In this paper, we assume that the region sizes of both BS and IRS are much smaller than the propagation distance, such that the far-field condition is satisfied and the plane-wave model can be used to form the field response \cite{MA3}. Let $\mathbf{G}\in\mathbb{C}^{N\times M}$ and $\mathbf{f}_k\in\mathbb{C}^{N\times 1}$ denote the BS-IRS channel and the channel between IRS and the $k$-th UE, respectively. Exploiting the field-response-based channel model \cite{MA7,MAchan}, the field response matrix (FRM) of the BS with $M$ MAs is
	\begin{equation}
		\mathbf{G}_t(\mathbf{t}) = \left[\mathbf{g}(t_1),\dots,\mathbf{g}(t_M)\right]\in\mathbb{C}^{L_t\times M},
	\end{equation}
	where $\mathbf{g}({t}_i)=\left[e^{\jmath\frac{2\pi}{\lambda}{\rho}_{t,1}t_i},\dots,e^{\jmath\frac{2\pi}{\lambda}{\rho}_{t,L_t}t_i}\right]^T\in\mathbb{C}^{L_t},\forall i\in\mathcal{B}$ is the field response vector (FRV) of an MA, $L_t$ is the number of transmit paths, ${\rho}_{t,l} = \cos \phi_{t,l}$, and ${\rho}_{t,l}{t}_i$ denote the difference of the signal propagation between position $t_i,\forall i\in\mathcal{B}$ and the origin of the transmit region for the $l$-th path, $\phi_{t,l}\in[0,\pi]$ is the angle of departure (AoD) of the $l$-th path.
	
	The FRM of the MA-IRS with the planar array for the arrival paths is defined as
	\begin{equation}
		\mathbf{F}_a(\mathbf{u}) = \left[\mathbf{f}_a(\mathbf{u}_1),\dots,\mathbf{f}_a(\mathbf{u}_N)\right]\in\mathbb{C}^{L_r\times N},
	\end{equation}
	where $\mathbf{f}_a(\mathbf{u}_i)=\left[e^{\jmath\frac{2\pi}{\lambda}\boldsymbol{\rho}_{r,1}^T\mathbf{u}_i},\dots,e^{\jmath\frac{2\pi}{\lambda}\boldsymbol{\rho}_{r,L_r}^T\mathbf{u}_i}\right]^T\in\mathbb{C}^{L_r},\forall i \in\mathcal{N}$ is the FRV of an MA of the MA-IRS, $L_r$ is the number of arriving paths, $\boldsymbol{\rho}_{r,l} = [\sin \theta_{r,l}\cos \phi_{r,l},\cos \theta_{r,l}]^T$, $\theta_{r,l}\in[0,\pi]$ and $\phi_{r,l}\in[0,\pi]$ are the elevation and azimuth angles of arrival (AoA) of the $l$-th path, respectively. We assume that the AoD and AoA of each IRS element are supplementary, then the FRM of the IRS for the reflecting paths is
	\begin{equation}
		\mathbf{F}_d(\mathbf{u}) = \left[\mathbf{f}_d(\mathbf{u}_1),\dots,\mathbf{f}_d(\mathbf{u}_N)\right]\in\mathbb{C}^{L_r\times N},
	\end{equation}
	where the FRV of each MA is calculated as $\mathbf{f}_d(\mathbf{u}_i)=\left[e^{-\jmath\frac{2\pi}{\lambda}\boldsymbol{\rho}_{r,1}^T\mathbf{u}_i},\dots,e^{-\jmath\frac{2\pi}{\lambda}\boldsymbol{\rho}_{r,L_r}^T\mathbf{u}_i}\right]^T\in\mathbb{C}^{L_r}, \forall i \in\mathcal{N}$.
	
	We consider the geometry model for the channel modeling \cite{MA7,MA9}, where the numbers of transmit and receive paths are the same ($L_t=L_r=L$). Define the path response matrix of the BS-IRS and IRS-UE channels as $\mathbf{\Sigma}_\mathrm{G} =\operatorname{diag}([\sigma_{G,1},\dots,\sigma_{G,L}]^T)\in \mathbb{C}^{L \times L}$ and $\mathbf{\Sigma}_{\mathrm{f},k} =\operatorname{diag}([\sigma_{k,1},\dots,\sigma_{k,L}]^T) \in \mathbb{C}^{L \times L}, \forall k \in \mathcal{K}$, where $\sigma_{G,l}$ and $\sigma_{k,l}$ are the complex response of the $l$-th path of the BS-IRS and IRS-UE channels, respectively. Based on the above, the BS-IRS channel matrix is given as 
	\begin{equation}
		\mathbf{G} = \mathbf{F}_a(\mathbf{u})^H\mathbf{\Sigma}_\mathrm{G}\mathbf{G}_t(\mathbf{t}),
	\end{equation}
	and the element of which can be calculated as
	\begin{equation}
		\mathbf{G}[n,m] =\sum_{l=1}^{L} \sigma_{G,l}e^{\jmath\frac{2\pi}{\lambda}({\rho}_{t,l}{t}_m-\boldsymbol{\rho}_{r,l}^T\mathbf{u}_n)},\forall m\in\mathcal{B},n\in\mathcal{N}.
	\end{equation}
	
	Similarly, the channel vector between IRS and the $k$-th UE is given by
	\begin{equation}
		\mathbf{f}_k = \mathbf{F}_d(\mathbf{u})^H\mathbf{\Sigma}_{\mathrm{f},k}\mathbf{1}
	\end{equation}
	with element value
	\begin{equation}
		\mathbf{f}_k[n] =\sum_{l=1}^{L} \sigma_{k,l}e^{\jmath\frac{2\pi}{\lambda}(\boldsymbol{\rho}_{r,l}^T\mathbf{u}_n)},\forall n\in\mathcal{N}.
	\end{equation}
	
	Denote the phase shift matrix of the IRS as $\mathbf{\Phi} = \operatorname{diag}(e^{j\phi_1},\dots,e^{j\phi_N})$, where $\phi_i \in [0,2\pi), \forall i\in{1,\dots,N}$ is the phase shift of each IRS element. To facilitate derivations, we define the phase shift vector of the IRS as $\boldsymbol{\phi} = [e^{j\phi_1},\dots, e^{j\phi_N}]^H\in\mathbb{C}^N$, then the equivalent channel between the BS and the $k$-th UE can be denoted as
	\begin{equation}
		\mathbf{h}^H_k = \mathbf{f}^H_{k}\mathbf{\Phi}\mathbf{G}=\boldsymbol{\phi}^H\operatorname{diag}(\mathbf{f}^H_{k})\mathbf{G}, \forall k \in \mathcal{K},
	\end{equation}
	which is a function with respect to the IRS phase shifts and the MAPs including $\mathbf{t}$ and $\mathbf{u}$. 
	
	It can be observed that both the BS-IRS and IRS-UE channels are characterized by the field response information (FRI) including the AoDs, AoAs, and the FRMs. To characterize the theoretical performance of the MA-IRS-aided communication systems, we assume that perfect FRI is known given the advanced channel estimation (CE) technologies for the MA-enabled communication systems \cite{MAchan,CE}. Nevertheless, we will show that the proposed method can achieve a robust performance with imperfect FRI via simulations in Section IV. Denote the transmit signal symbols for the $k$-th UE as $s_k\in\mathbb{C}$, the received signal at the $k$-th UE is  
	\begin{flalign}
		y_k = \mathbf{h}_k^H\mathbf{w}_k s_k + \!\!\sum_{j\in\mathcal{K},j\neq k}\!\!\mathbf{h}_k^H\mathbf{w}_j s_j + n_k,
	\end{flalign}
	where $\mathbf{w}_k$ is the precoding vector for the $k$-th UE, $n_k\sim{\mathcal{CN}(0,\sigma_n^2)}$ is the additive complex Gaussian noise. Denote the precoding matrix as $\mathbf{W}=[\mathbf{w}_1,\dots,\mathbf{w}_k]\in\mathbb{C}^{M\times K}$ with the maximum transmit power $P_t$, then the achievable transmission rate of the $k$-th UE is obtained as $r_k = \log\bigl(1+\gamma_k\bigr)$, where
	\begin{flalign}
		\gamma_k=\frac{\lvert\mathbf{h}_k^H\mathbf{w}_k\rvert^2}{\sum_{j\in\mathcal{K},j\neq k}\lvert\mathbf{h}_k^H\mathbf{w}_j\rvert^2 + \sigma_n^2}
	\end{flalign} 
	is the SINR for the $k$-th UE. Since the BS and IRS are static, we assume that they are connected to a controller via wired links and can be configured simultaneously exploiting the given FRI.
	
	\subsection{Problem Formulation}
	For such a multi-user communication system, a commonly adopted performance metric is the sum-rate, which indicates the total throughput of the system and can be given by $\sum_{k\in\mathcal{K}}r_k$. However, the performance obtained by each individual UE is not guaranteed while maximizing the sum-rate. This may result in unfair power allocation and significantly lower performance for some users \cite{IRS6,IRS7}. To avoid the unfairness issue and ensure the QoS of minimum rate obtained by the UEs, we aim to maximize the sum-rate of the system while ensuring a rate threshold by jointly optimizing the variables including the precoding matrix $\mathbf{W}$, the IRS phase shift vector $\boldsymbol{\phi}$, and the MAPs including $\mathbf{t}$ and $\mathbf{u}$. Denote the minimum rate threshold as $\Gamma$, the problem is formulated as
	\begin{subequations}
		\label{Qsr}
		\begin{IEEEeqnarray}{r,l}			
			$$\underset{\mathbf{W},\boldsymbol{\phi}, \mathbf{t},\mathbf{u}}{\min}$$&{\ -\sum_{k\in\mathcal{K}} r_k}\\
			$$\operatorname{s.t.}$$ &\ \operatorname{Tr}(\mathbf{W}\mathbf{W}^H) \leq P_t,  \label{Cp}\\
			&\ \lvert \phi_{n}\rvert=1, \ \forall n\in\mathcal{N}, \label{Cphi}\\
			&\ r_k \geq \Gamma, \forall k \in \mathcal{K}, \label{Cr}\\
			&\ {t}_m \in \mathcal{C}_B, \forall m \in \mathcal{B} \label{Cbregion}\\
			&\ \mathbf{u}_n \in \mathcal{C}_I, \forall n\in\mathcal{N} \label{Ciregion}\\
			&\ \lvert {t}_m-{t}_{m'}\rvert \geq \lambda/2, \forall m\neq m', m,m'\in\mathcal{B} \label{Ct}\\
			&\ \lVert \mathbf{u}_n-\mathbf{u}_{n'}\rVert \geq \lambda/2, \forall n\neq n',n,n'\in\mathcal{N}. \label{Cu}
		\end{IEEEeqnarray}
	\end{subequations}
	
	It can be observed that problem \eqref{Qsr} is highly non-convex due to the non-convex objective functions and constraints. Existing AO-based methods for conventional FPA-IRS-aided communications cannot be utilized directly to solve the problem well. Specifically, existing precoding techniques such as ZF and WMMSE cannot be exploited directly since \eqref{Cr} cannot be guaranteed, and the MAPs are coupled with the channel matrices and hard to tackle. In the following sections, we propose a PRMO method to solve the problem, where all the variables, including the precoding matrix, the IRS phase shift vector, and the MAPs are optimized simultaneously to obtain a feasible solution.
	
	\section{PRMO For The Joint Beamforming and MAP Optimization}
	In this section, we develop the PRMO method to solve \eqref{Qsr}. The main steps are as follows: 1) A PRMS is constructed as the solution space and the problem is reformulated over the PRMS; 2) The problem is transformed via penalty method and smoothing technique to tackle the minimum rate constraints \eqref{Cr} and the minimum MA distance constraints \eqref{Ct} and \eqref{Cu}; 3) The RBFGS algorithm is exploited for updating the variables over the PRMS; 4) The REP method is utilized to find a feasible solution of \eqref{Qsr}.
	
	\subsection{Construction of the PRMS}
	It is intuitive that the transmit power should be fully utilized to maximize the sum-rate, thus the constraint \eqref{Cp} can be restricted as $\operatorname{Tr}(\mathbf{W}\mathbf{W}^H) = P_t$, which can be interpreted as restricting the variable over the Riemannian complex sphere manifold (CSM) \cite{CSM}. The CSM is given by
	\begin{equation}
		\mathcal{M}_{\mathbf{W}}=\{\mathbf{W}\in \mathbb{C}^{N\times K} \mid \operatorname{Tr}(\mathbf{W}\mathbf{W}^H) = P_t\}.
	\end{equation}
	
	Besides, the constant modulus constraints (CMC) of the IRS phase shift vector \eqref{Cphi} can be interpreted as constraining the variable to the Riemannian complex circle manifold (CCM) \cite{rmo}, which is given by
	\begin{flalign}	
		\mathcal{M}_{\boldsymbol{\phi}}=\{\boldsymbol{\phi}\in \mathbb{C}^{N} \mid \lvert \phi_n\rvert=1, \forall n \in \mathcal{N}\}.
	\end{flalign} 	
	Note that the CSM and CCM can be linearized locally around every point as tangent spaces. The tangent spaces of a point over the CSM and CCM are given as
	\begin{flalign}	
		{\rm T}_\mathbf{W}\mathcal{M}_\mathbf{W} \!=\!\bigl\{\boldsymbol{\zeta}_\mathbf{W}\!\mid\! \boldsymbol{\zeta}_\mathbf{W}\!\in\!\mathbb{C}^{M\times K}, \Re\{\operatorname{Tr}(\mathbf{W}^H\boldsymbol{\zeta}_\mathbf{W})\}\!=\!0\bigr\}
	\end{flalign}
	and
	\begin{flalign}	
		{\rm T}_{\boldsymbol{\phi}}\mathcal{M}_{\boldsymbol{\phi}} \!=\!\bigl\{\boldsymbol{\zeta}_{\boldsymbol{\phi}}\!\mid\! \boldsymbol{\zeta}_{\boldsymbol{\phi}} \in \mathbb{C}^{N}, \Re(\boldsymbol{\zeta}^*_{\boldsymbol{\phi}}\odot\boldsymbol{\phi})=\mathbf{0}\bigr\}.
	\end{flalign}
	
	To satisfy the constraints \eqref{Cbregion} and \eqref{Ciregion} that confine the MAPs in the given region, we introduce auxiliary variables $\mathbf{o}=[{o}_1,\dots,{o}_M]^T\in\mathbb{R}^{M}$ and $\mathbf{p}=[\mathbf{p}^T_1,\dots,\mathbf{p}^T_N]^T=[p_1,\dots,p_{2N}]^T\in\mathbb{R}^{2N}$. Then, exploiting the $\tanh$ function, which is defined as $\tanh(x)=(e^x-e^{-x})/(e^x+e^{-x})\in(-1,1)$, we define the projection functions on $\mathbf{o}$ and $\mathbf{p}$ as
	\begin{equation}
		\boldsymbol{p}_B(\mathbf{o}) = \frac{A_{B}}{2}\boldsymbol{\tanh}(\mathbf{o}),\ 
		\boldsymbol{p}_I(\mathbf{p}) = \frac{A_{I}}{2}\boldsymbol{\tanh}(\mathbf{p}),
	\end{equation}
	where $\boldsymbol{\tanh}(\mathbf{o})=[\tanh(o_1),\dots,\tanh(o_M)]^T\in\mathbb{R}^{M}$ and $\boldsymbol{\tanh}(\mathbf{p})=[\tanh(p_1),\dots,\tanh(p_{2N})]^T\in\mathbb{R}^{2N}$. Hence, problem \eqref{Qsr} is equivalent to 
	\begin{subequations}
		\label{Qsr2}
		\begin{IEEEeqnarray}{r,l}			
			$$\underset{\mathbf{W},\mathbf{o},\mathbf{p}}{\min}$$&{\ -\sum_{k\in\mathcal{K}} r_k}\\
			$$\operatorname{s.t.}$$ &\ \mathbf{W} \in \mathcal{M}_{\mathbf{W}},\\
			&\ \boldsymbol{\phi} \in \mathcal{M}_{\boldsymbol{\phi}},\\
			&\ r_k \geq \Gamma, \forall k \in \mathcal{K},\\
			&\ \mathbf{o}\in\mathbb{R}^{M},\mathbf{p}\in\mathbb{R}^{2N},\\
			&\ \mathbf{t}= \boldsymbol{p}_B(\mathbf{o}),\mathbf{u} = \boldsymbol{p}_I(\mathbf{p}), \\
			&\ \lVert {t}_m-{t}_{m'}\rVert \geq \lambda/2, \forall m\neq m', m,m'\in\mathcal{B} \\
			&\ \lVert \mathbf{u}_n-\mathbf{u}_{n'}\rVert \geq \lambda/2, \forall n\neq n', n,n'\in\mathcal{N}.		
		\end{IEEEeqnarray}
	\end{subequations}
	
	Define the variable $\mathbf{X} = \left[\mathbf{W},\boldsymbol{\phi},\mathbf{o},\mathbf{p}\right]$. Note that $\mathbf{o}$ and $\mathbf{p}$ are confined in Euclidean real space, which is also a basic RMS \cite{rmo}. Combining the RMSs of the variables, a PRMS $\mathcal{M}$ can be constructed as the solution space for $\mathbf{X}$, which is given as
	\begin{align}	
		\mathcal{M} =\{\mathbf{X}&=(\mathbf{W},\boldsymbol{\phi},\mathbf{o},\mathbf{p}) \mid \nonumber\\&\mathbf{W}\in\mathcal{M}_{\mathbf{W}}, \boldsymbol{\phi}\in\mathcal{M}_{\boldsymbol{\phi}},\mathbf{o}\in\mathbb{R}^{ M},\mathbf{p}\in\mathbb{R}^{2N}\}.
	\end{align}
	The tangent space of the point $\mathbf{X}$ over $\mathcal{M}$ is then obtained as
	\begin{align}	
		{\rm T}_{\mathbf{X}} \mathcal{M} =\bigl\{\boldsymbol{\zeta}_{\mathbf{X}}& = (\boldsymbol{\zeta}_\mathbf{W},\boldsymbol{\zeta}_{\boldsymbol{\phi}},\boldsymbol{\zeta}_\mathbf{o},\boldsymbol{\zeta}_\mathbf{p})\mid \boldsymbol{\zeta}_\mathbf{W}\in {\rm T}_\mathbf{W}\mathcal{M}_\mathbf{W},\nonumber\\&\boldsymbol{\zeta}_{\boldsymbol{\phi}}\in {\rm T}_{\boldsymbol{\phi}}\mathcal{M}_{\boldsymbol{\phi}},\boldsymbol{\zeta}_\mathbf{o}\in\mathbb{R}^{M},\boldsymbol{\zeta}_\mathbf{p}\in\mathbb{R}^{2N}\bigr\}.
	\end{align}
	\\ 
	We equip $\mathcal{M}$ with the Euclidean inner product as the Riemannian metric to obtain the inner product over $\mathcal{M}$, which is defined as
	\begin{align}	&\langle\boldsymbol{\zeta}_{\mathbf{X}},\boldsymbol{\zeta}'_{\mathbf{X}}\rangle=\langle(\boldsymbol{\zeta}_\mathbf{W},\boldsymbol{\zeta}_{\boldsymbol{\phi}},\boldsymbol{\zeta}_\mathbf{o},\boldsymbol{\zeta}_\mathbf{p}),(\boldsymbol{\zeta}'_\mathbf{W},\boldsymbol{\zeta}'_{\boldsymbol{\phi}},\boldsymbol{\zeta}'_\mathbf{o},\boldsymbol{\zeta}'_\mathbf{p})\rangle\nonumber\\ &=\Re\left(\operatorname{Tr}(\boldsymbol{\zeta}^H_{\mathbf{W}}\boldsymbol{\zeta}'_{\mathbf{W}})+\boldsymbol{\zeta}^H_{\boldsymbol{\phi}}\boldsymbol{\zeta}'_{\boldsymbol{\phi}}\right)  +\boldsymbol{\zeta}_\mathbf{o}^T\boldsymbol{\zeta}_\mathbf{o}+ \boldsymbol{\zeta}_\mathbf{p}^T\boldsymbol{\zeta}_\mathbf{p},
	\end{align}
	where $\boldsymbol{\zeta}_{\mathbf{X}}, \boldsymbol{\zeta}'_{\mathbf{X}} \in  {\rm T}_\mathbf{X}\mathcal{M}$, $\boldsymbol{\zeta}_{\mathbf{W}}, \boldsymbol{\zeta}'_{\mathbf{W}} \in  {\rm T}_\mathbf{W}\mathcal{M}_\mathbf{W}$, $\boldsymbol{\zeta}_{\boldsymbol{\phi}},\boldsymbol{\zeta}'_{\boldsymbol{\phi}}\in{\rm T}_{\boldsymbol{\phi}} \mathcal{M}_{\boldsymbol{\phi}}$, $\boldsymbol{\zeta}_{\mathbf{o}}, \boldsymbol{\zeta}'_{\mathbf{o}} \in  \mathbb{R}^{M}$, and $\boldsymbol{\zeta}_{\mathbf{p}}, \boldsymbol{\zeta}'_{\mathbf{p}} \in  \mathbb{R}^{2N}$. The norm over $\mathcal{M}$ can be then calculated as $\lVert \boldsymbol{\zeta}_\mathbf{X} \rVert = \sqrt{\langle \boldsymbol{\zeta}_{\mathbf{X}},\boldsymbol{\zeta}_{\mathbf{X}}\rangle}$.

	\subsection{Problem Reformulation via Penalty Method}
	The inequality constraints of minimum MAP distance \eqref{Ct} and \eqref{Cu} can be rewritten as
	\begin{align}
		h_{m,m'}(\mathbf{X})=\lambda/2 - \lvert p_B({o}_m) - &p_B({o}_{m'}) \rvert \leq 0,\nonumber\\ &\forall m,m' \in \mathcal{B}, m \neq m'
	\end{align}
	and 
	\begin{align}
		h_{n,n'}(\mathbf{X})=\lambda/2 - \lVert p_I(\mathbf{p}_n) - &p_I(\mathbf{p}_{n'}) \rVert \leq 0, \nonumber\\&\forall n,n' \in \mathcal{N}, n \neq n'.
	\end{align} 
	
	We define the set of the minimum MA distance constraints as $\mathcal{D}_B = \{(i,i')\mid i,i'\in\mathcal{B}, i\neq i'\}$ and $\mathcal{D}_I = \{(i,i')\mid \ i,i'\in\mathcal{N}, i\neq i'\}$. Besides, the minimum rate threshold constraints can be rewritten as 
	\begin{equation}
		h_k(\mathbf{X}) = \Gamma - r_k \leq 0,\forall k \in \mathcal{K}.
	\end{equation}
	
	Denote the set of all the inequality constraints as $\mathcal{I} = \mathcal{D}_B \cup \mathcal{D}_I \cup \mathcal{K}$, the problem \eqref{Qsr2} can be rewritten as
	\begin{subequations}
		\label{Qsr3}
		\begin{IEEEeqnarray}{r,l}			
			$$\underset{\mathbf{X}}{\min}$$&{\ f(\mathbf{X})=-\sum_{k\in\mathcal{K}} r_k}\\
			$$\operatorname{s.t.}$$ &\ \mathbf{X} \in \mathcal{M},\\
			&\ h_i(\mathbf{X}) \leq 0, \forall i\in\mathcal{I}. \label{ineq}
		\end{IEEEeqnarray}
	\end{subequations}
	
	To tackle the inequality constraint \eqref{ineq}, a commonly adopted approach is the penalty method, which supplements the objective function with a weighted penalty $\rho \sum_{i\in\mathcal{I}}\max\{0,h_i(\mathbf{\mathbf{X}})\}$ for violating the constraint, where $\rho\geq0$ is the penalty weight \cite{penalty, ep}. In the Euclidean case, it is known that exact satisfaction of constraints can be achieved by a finite penalty weight \cite{penalty}, and it can be extended in the Riemannian case \cite{rep}. However, the penalty is intractable since it is nonsmooth and has non-differential discontinuities. To tackle the two-term maximization, the log-sum-exp (LSE) function is a commonly adopted smoothing technique, which is continuous differential and convex. LSE is given by $\max\{a,b\}\approx u\log(e^{a/u}+e^{b/u})$, where $u\geq0$ is the smoothing parameter. Typically, lower $u$ leads to higher approximation accuracy. Exploiting the weighted penalty and LSE, \eqref{Qsr3} can be converted as
	\begin{subequations}
		\label{Qsrs}
		\begin{IEEEeqnarray}{r,l}			
			$$\underset{\mathbf{X}}{\min}$$&{\  g(\mathbf{X})=f(\mathbf{X})+\rho\!\sum_{i\in \mathcal{I}}\!u\log(1+e^{h_i(\mathbf{X})/u})} \\
			$$\operatorname{s.t.}$$ &\ \mathbf{X}\in\mathcal{M}.
		\end{IEEEeqnarray}
	\end{subequations}
	
	With proper values of $\rho$ and $u$, a feasible solution of \eqref{Qsr3} can be found by finding an optimum of \eqref{Qsrs}. However, the optimal values of $\rho$ and $u$ are difficult to find. Moreover, while \eqref{Qsrs} is an unconstrained smooth problem over $\mathcal{M}$, it is non-convex over $\mathcal{M}$ and hard to solve. In the next subsections, we first exploit the RBFGS algorithm over $\mathcal{M}$ to solve \eqref{Qsrs} with fixed $\rho$ and $u$, then we utilize REP to find proper values of $\rho$ and $u$ that ensure the inequality constraints.
	
	\subsection{RBFGS Algorithm for the Sub-problem}
	
	To solve \eqref{Qsrs} with fixed $\rho$ and $u$ by updating the variable over $\mathcal{M}$, we propose the RBFGS algorithm to calculate update directions exploiting the derived Riemannian gradient and obtain a feasible optima in this subsection. 
	
	\subsubsection{Obtain the Riemannian gradient over $\mathcal{M}$}
	First, for the precoding matrix $\mathbf{W}$ and IRS phase shift vector, we obtain the Euclidean gradients 
	\begin{equation}
		\nabla_{\mathbf{W}^*} g(\mathbf{X}) = \nabla_{\mathbf{W}^*} f({\mathbf{X}}) + \rho \sum_{i\in\mathcal{K}}\lambda_i \nabla_{\mathbf{W}^*}h_i(\mathbf{X}),
	\end{equation}
	and
	\begin{equation}
		\nabla_{\boldsymbol{\phi}^*} g(\mathbf{X}) = \nabla_{\boldsymbol{\phi}^*} f({\mathbf{X}}) + \rho \sum_{i\in\mathcal{K}}\lambda_i \nabla_{\boldsymbol{\phi}^*}h_i(\mathbf{X}),
	\end{equation}
	where
	\begin{equation}
		\nabla_{\mathbf{W}^*}f(\mathbf{X})= -\sum_{k\in\mathcal{K}}\frac{\partial r_k}{\partial {\mathbf{W}^*}}, \nabla_{\boldsymbol{\phi}^*}f(\mathbf{X})= -\sum_{k\in\mathcal{K}}\frac{\partial r_k}{\partial {\boldsymbol{\phi}^*}},
	\end{equation}
	\begin{equation}
		\nabla_{\mathbf{W}^*}h_i(\mathbf{X})= -\frac{\partial r_i}{\partial {\mathbf{W}^*}}, \nabla_{\boldsymbol{\phi}^*}h_i(\mathbf{X})= -\frac{\partial r_i}{\partial {\boldsymbol{\phi}^*}}, \forall i \in \mathcal{K},
	\end{equation}
	and
	\begin{flalign}	
		\lambda_i = \frac{e^{h_i(\mathbf{X})/u}}{1+e^{h_i(\mathbf{X})/u}}.
	\end{flalign}
	
	Similarly, we can calculate the Euclidean gradients with respect to the MAPs. Take $\mathbf{o}$ as an example, we can obtain 
	\begin{equation}
		\nabla_\mathbf{o} g(\mathbf{X}) = \nabla_{\mathbf{o}}f({\mathbf{X}}) + \rho \sum_{i\in\mathcal{I}}\lambda_i \nabla_{\mathbf{o}}h_i(\mathbf{X}),
	\end{equation}
	where 
	\begin{equation}
		\nabla_{\mathbf{o}}f(\mathbf{X}) =-\sum_{k\in\mathcal{K}}\frac{\partial r_k}{\partial \boldsymbol{p}_B(\mathbf{o})}\odot \frac{A_B}{2}(\mathbf{1}-\boldsymbol{\tanh}^2(\mathbf{o}))
	\end{equation}
	and
	\begin{equation}
		\nabla_{\mathbf{o}}h_i(\mathbf{X}) =\frac{\partial h_i(\mathbf{X})}{\partial \boldsymbol{p}_B(\mathbf{o})}\odot \frac{A_B}{2}(\mathbf{1}-\boldsymbol{\tanh}^2(\mathbf{o})),\forall i \in \mathcal{D}_B.
	\end{equation}
	
	The Euclidean gradients $\nabla_\mathbf{p} g(\mathbf{X})$ can be calculated using the same way. The detailed partial derivatives are provided in Appendix A. Then, the Euclidean gradient of $g(\mathbf{X})$ with respect to $\mathbf{X}$ is given as 
	\begin{align}
		\nabla_{\mathbf{X}}g(\mathbf{X})=&[\nabla_{\mathbf{W}}g(\mathbf{X}),\nabla_{\boldsymbol{\phi}}g(\mathbf{X}),\nabla_{\mathbf{o}}g(\mathbf{X}),\nabla_{\mathbf{p}}g(\mathbf{X})]
		\nonumber\\=& \nabla_{\mathbf{X}}f({\mathbf{X}}) + \rho \sum_{i\in\mathcal{I}}\lambda_i \nabla_{\mathbf{X}}h_i(\mathbf{X}).
	\end{align}
	
	The Riemannian gradient is calculated as the orthogonal projection of the Euclidean gradient to the tangent space of a point over $\mathcal{M}$. Specifically, the Riemannian gradient with respect to $\mathbf{W}$ and $\boldsymbol{\phi}$ are given by
	\begin{align}	
		\operatorname{grad}_{\mathbf{W}}g(\mathbf{X}) =\nabla_{\mathbf{W}^*}g(\mathbf{X})- \mathbf{W}\Re\{\operatorname{Tr}\bigl(\mathbf{W}^H\nabla_{\mathbf{W}^*}g(\mathbf{X})\bigr)\}
	\end{align}
	and
	\begin{align}	
		\operatorname{grad}_{\boldsymbol{\phi}}g(\mathbf{X}) =\nabla_{\boldsymbol{\phi}^*}g(\mathbf{X})- \Re\{\nabla_{\boldsymbol{\phi}^*} g(\mathbf{X})\odot\boldsymbol{\phi}^*\}\odot \boldsymbol{\phi}
	\end{align}
	
	Since $\mathbf{o}$ and $\mathbf{p}$ are constrained in the RMSs of Euclidean real space, the values of the corresponding Riemannian gradients are equal to the Euclidean gradients, e.g., $\operatorname{grad}_\mathbf{o}g(\mathbf{X})=\nabla_{\mathbf{o}}g(\mathbf{X})$. Then, the Riemannian gradient with respect to $\mathbf{X}$ is obtained as
	\begin{align}
		\label{rg}
		&\operatorname{grad}_{\mathbf{X}}g(\mathbf{X})\nonumber\\ =&\big[\operatorname{grad}_{\mathbf{W}}g(\mathbf{X}),\operatorname{grad}_{\boldsymbol{\phi}}g(\mathbf{X}), \operatorname{grad}_{\mathbf{o}}g(\mathbf{X}),\operatorname{grad}_{\mathbf{p}}g(\mathbf{X})\big]
		\nonumber\\= &\operatorname{grad}_{\mathbf{X}}f(\mathbf{X}) + \rho \sum_{i\in\mathcal{I}}\lambda_i \operatorname{grad}_{\mathbf{X}}h_i(\mathbf{X}).
	\end{align}
	
	\subsubsection{RBFGS algorithm for obtaining the update direction}
	Although the first-order gradient descent (GD) and gradient projection (GP) methods can be applied directly to find a solution exploiting the derived gradients \cite{MA7,MAIRS1}, they can be vulnerable to the local optima in such a highly non-convex solution space. Instead, the second-order method exploiting curvature information can be more effective for such a problem. For the Riemannian case, an optimal second-order descent direction $\mathbf{d}_\mathbf{X}$ of $g(\mathbf{X})$ can be obtained via the Newton equation as $\operatorname{Hess}_\mathbf{X}g(\mathbf{x})\mathbf{d}_\mathbf{X}=-\operatorname{grad}_{\mathbf{X}}g(\mathbf{X})$, where $\operatorname{Hess}(\cdot)$ denotes the Hessian matrix \cite{rmo}. However, it is expensive to calculate the inverse Hessian matrix, and the Hessian may be not positive definite. 
	
	Instead, we resort to the quasi-Newton method of BFGS algorithm to obtain an approximation of the Hessian inverse for calculating the update direction \cite{penalty}. By extending the BFGS direction in the Euclidean case to the Riemannian case \cite{penalty,rbfgs}, the update direction over the constructed PRMS is defined as 
	\begin{align}	
		\mathbf{d}_{\mathbf{X}}&= (\mathbf{d}_\mathbf{W},\mathbf{d}_{\boldsymbol{\phi}},\mathbf{d}_\mathbf{o},\mathbf{d}_\mathbf{p}) = -\mathbf{H}_{\mathbf{X}}\operatorname{grad}_\mathbf{X}g(\mathbf{X})\nonumber\\& =- \bigl(\mathbf{H}_\mathbf{W}\operatorname{grad}_{\mathbf{W}}  g(\mathbf{X}),\mathbf{H}_{\boldsymbol{\phi}}\operatorname{grad}_{\boldsymbol{\phi}}  g(\mathbf{X}),\nonumber\\&\qquad\ \ \  \mathbf{H}_\mathbf{o}\operatorname{grad}_{\mathbf{o}}  g(\mathbf{X}),\mathbf{H}_\mathbf{p}\operatorname{grad}_{\mathbf{p}}  g(\mathbf{X})\bigr),
	\end{align}
	where $\mathbf{H}_{\mathbf{X}}$ is the approximation to the inverse Hessian. 
	
	To calculate the Hessian inverse approximations over $\mathcal{M}$, we first define the following operations. The retraction operation for updating the variable $\mathbf{X}$ over $\mathcal{M}$ with an update direction $\mathbf{d}_\mathbf{X}$ and step size $\alpha$ on $\mathcal{M}$ is defined as
	\begin{align}
		\label{ret}
		\mathcal{R}_{\mathbf{X}}&(\alpha \mathbf{d}_{\mathbf{X}}) =\bigl( \sqrt{{{P_t}}/{\lVert\mathbf{W}+\alpha\mathbf{d}{\mathbf{W}}\rVert}}(\mathbf{W}+\alpha\mathbf{d}_{\mathbf{W}}),\nonumber\\&(\boldsymbol{\phi}+\alpha\mathbf{d}_{\boldsymbol{\phi}}) \oslash \lvert\boldsymbol{\phi}+\alpha\mathbf{d}_{\boldsymbol{\phi}}\rvert,  \mathbf{o}+\alpha\mathbf{d}_{\mathbf{o}},\mathbf{p}+\alpha\mathbf{d}_{\mathbf{p}} \bigr).
	\end{align}
	
	Since the calculation involving the vectors in different tangent spaces, the transportation operation that relocates an arbitrary direction to the tangent space of a point $\mathbf{X}$ is defined as
	\begin{align}
		\label{tra}
		\mathcal{T}_{\mathbf{X}}(\mathbf{d})= \bigl(&\mathbf{d}_\mathbf{W} - \mathbf{W}\Re\{\operatorname{Tr}(\mathbf{W}^H\mathbf{d}_\mathbf{W})\},\nonumber\\&\ \  \mathbf{d}_{\boldsymbol{\phi}} - \Re\{\mathbf{d}_{\boldsymbol{\phi}}^*\odot\boldsymbol{\phi}'\}\odot \boldsymbol{\phi}', \mathbf{d}_\mathbf{o},\mathbf{d}_\mathbf{p} \bigr).
	\end{align}
	
	By extending the BFGS algorithm in Euclidean case to the PRMS \cite{penalty,rbfgs}, the inverse Hessian approximation $\mathbf{H}_\mathbf{X}$ of the $(l+1)$-th iteration can be obtained as follows. First, we define the following medium variables 
	\begin{equation}
		\label{s}
		\mathbf{s}^l_\mathbf{X} = (\mathbf{s}^l_\mathbf{W},\mathbf{s}^l_{\boldsymbol{\phi}},\mathbf{s}^l_\mathbf{o},\mathbf{s}^l_\mathbf{p})= \mathcal{T}_{\mathbf{X}^{l+1}}(\alpha^l\mathbf{d}_{\mathbf{X}}^l)
	\end{equation}
	and
	\begin{align}
		\label{y}
		\mathbf{y}^l_\mathbf{X}& =(\mathbf{y}^l_\mathbf{W},\mathbf{y}^l_{\boldsymbol{\phi}},\mathbf{y}^l_\mathbf{o},\mathbf{y}^l_\mathbf{p})\nonumber\\&= \operatorname{grad}_{\mathbf{X}}g(\mathbf{X}^{l+1}) - \mathcal{T}_{\mathbf{X}^{l+1}}(\operatorname{grad}_{\mathbf{X}}g(\mathbf{X}^{l})).
	\end{align}
	
	To avoid the scaling of the RBFGS direction, the medium variables should be normalized, which can be implemented as $\mathbf{s}^l = \mathbf{s}_{\mathbf{X}}^l/\lVert\mathbf{s}_{\mathbf{X}}^l\rVert$ and $\mathbf{y}_{\mathbf{X}}^l = \mathbf{y}_{\mathbf{X}}^l/\lVert\mathbf{s}_{\mathbf{X}}^l\rVert$. Then, the inverse Hessian approximation with respect to $\mathbf{W}$, $\boldsymbol{\phi}$, $\mathbf{o}$, and $\mathbf{p}$ can be obtained as
	\begin{align}
		\label{Hw}
		\mathbf{H}^{l+1}_\mathbf{W} =(\mathbf{V}_{\mathbf{W}} ^l)^H\mathbf{H}_\mathbf{W}^{l}\mathbf{V}_{\mathbf{W}}^l + \delta^l\mathbf{s}^l_{\mathbf{W}}(\mathbf{s}^l_{\mathbf{W}})^H,
	\end{align}
	\begin{align}
		\label{Hphi}
		\mathbf{H}^{l+1}_{\boldsymbol{\phi}} =(\mathbf{V}_{\boldsymbol{\phi}} ^l)^H\mathbf{H}_{\boldsymbol{\phi}}^{l}\mathbf{V}_{\boldsymbol{\phi}}^l + \delta^l\mathbf{s}^l_{{\boldsymbol{\phi}}}(\mathbf{s}^l_{{\boldsymbol{\phi}}})^H,
	\end{align}
	\begin{align}
		\label{Ho}
		\mathbf{H}^{l+1}_\mathbf{o} =(\mathbf{V}_{\mathbf{o}} ^l)^H\mathbf{H}_\mathbf{o}^{l}\mathbf{V}_{\mathbf{o}}^l + \delta^l\mathbf{s}^l_{\mathbf{o}}(\mathbf{s}^l_{\mathbf{o}})^H,
	\end{align}
	and
	\begin{align}
		\label{Hp}
		\mathbf{H}^{l+1}_\mathbf{p} =(\mathbf{V}_{\mathbf{p}} ^l)^H\mathbf{H}_\mathbf{p}^{l}\mathbf{V}_{\mathbf{p}}^l + \delta^l\mathbf{s}^l_{\mathbf{p}}(\mathbf{s}^l_{\mathbf{p}})^H,
	\end{align}
	
	where $\delta^l=1/\langle\mathbf{s}^l_{\mathbf{X}},\mathbf{y}^l_{\mathbf{X}}\rangle$, $\mathbf{V}_{\mathbf{W}} ^l=\mathbf{I}-\delta^l\mathbf{s}^l_{\mathbf{W}}(\mathbf{y}^l_{\mathbf{W}})^H$,  $\mathbf{V}_{\boldsymbol{\phi}} ^l=\mathbf{I}-\delta^l\mathbf{s}^l_{\boldsymbol{\phi}}(\mathbf{y}^l_{\boldsymbol{\phi}})^H$, $\mathbf{V}_{\mathbf{o}} ^l=\mathbf{I}-\delta^l\mathbf{s}^l_{\mathbf{o}}(\mathbf{y}^l_{\mathbf{o}})^H$, and  $\mathbf{V}_{\mathbf{p}} ^l=\mathbf{I}-\delta^l\mathbf{s}^l_{\mathbf{p}}(\mathbf{y}^l_{\mathbf{p}})^H$. Combining the above, $\mathbf{H}^{l+1}_{\mathbf{X}}$ can be obtained as
	\begin{equation}
		\label{H1}
		\mathbf{H}^{l+1}_{\mathbf{X}}=\left[\mathbf{H}^{l+1}_\mathbf{W},\mathbf{H}^{l+1}_{\boldsymbol{\phi}},\mathbf{H}^{l+1}_\mathbf{o},\mathbf{H}^{l+1}_\mathbf{p}\right].
	\end{equation}
	\begin{figure*}[ht]		
		\begin{align}			
			\label{H}
			\mathbf{H}^{l+1}_\mathbf{W}&=\bigl(\mathbf{V}^{l-m}_\mathbf{W}\dots\mathbf{V}^{l-1}_\mathbf{W}\bigr)^H \mathbf{H}^{l-m}_\mathbf{W}\bigl(\mathbf{V}^{l-m}_\mathbf{W}\dots\mathbf{V}^{l-1}_\mathbf{W}\bigr) + \delta^{l-m}\bigl(\mathbf{V}^{l-m+1}_\mathbf{W}\dots\mathbf{V}^{l-1}_\mathbf{W}\bigr)^H\mathbf{s}^{l-m}_\mathbf{W}(\mathbf{s}^{l-m}_\mathbf{W})^H\bigl(\mathbf{V}^{l-m+1}_\mathbf{W}\dots\mathbf{V}^{l-1}_\mathbf{W}\bigr)\nonumber\\&+\delta^{l-m+1}\bigl(\mathbf{V}^{l-m+2}_\mathbf{W}\dots\mathbf{V}^{l-1}_\mathbf{W}\bigr)^H\mathbf{s}^{l-m+1}_\mathbf{W}(\mathbf{s}^{l-m+1}_\mathbf{W})^H\bigl(\mathbf{V}^{l-m+2}_\mathbf{W}\dots\mathbf{V}^{l-1}_\mathbf{W}\bigr)+\dots+\delta^{l-1}\mathbf{s}^{l-1}_\mathbf{W}(\mathbf{s}^{l-1}_\mathbf{W})^H
		\end{align}	
		\hrulefill				
	\end{figure*}
	\subsubsection{RBFGS algorithm with limited memory}
	Although the Hessian is not calculated in the RBFGS algorithm, the inverse Hessian approximation $\mathbf{H}_\mathbf{X}$ calculated by \eqref{H1} will generally be dense, which leads to higher storing and computation cost, especially when the dimension of the variable is large. Instead, a modified version of it is utilizing limited memory. With a limited memory $\mathbf{M}$ with a size of $M_m$, a collection of no more than $M_m$ medium variable sets $\mathbf{M}_i = (\mathbf{s}^i_{\mathbf{X}},\mathbf{y}^i_{\mathbf{X}},\delta^i), i=1,\dots,M_m$ obtained in the last iterations can be stored and used to compute the update direction. Observing that \eqref{Hw}-\eqref{Hp} have recursive properties, take $\mathbf{H}^{l+1}_{\mathbf{W}}$ as an example, with $m\leq M_m$ stored memory, it can be expanded as \eqref{H}, as shown at the top of next page. By substituting $\delta^l$, $\mathbf{V}_{\mathbf{W}} ^l$, $\mathbf{V}_{\boldsymbol{\phi}} ^l$, $\mathbf{V}_{\mathbf{o}} ^l$, and $\mathbf{V}_{\mathbf{p}}^l$ into the recursive equations, $\mathbf{H}^{l+1}_{\mathbf{X}}$ can be calculated by a two-loop recursive procedure \cite{penalty,lbfgs,lrbfgs}. The computation process is summarized as \textbf{Algorithm 1}.

	\begin{algorithm}[t]
		\caption{RBFGS algorithm with limited memory for the $l$-th iteration}
		\begin{algorithmic}[1]
			\Require Initial direction $\mathbf{d}^l = \operatorname{grad} g(\mathbf{X}^l)$, $m$ stored medium variables $\mathbf{M}_i = (\mathbf{s}^i_{\mathbf{X}},\mathbf{y}^i_{\mathbf{X}},\delta^i), i=1,\dots,m$. 
			\For{$i= m : -1 : 1$}
			\State $\varrho^i= \delta^i \langle \mathbf{s}^i , \mathbf{d}^l\rangle$;
			\State $\mathbf{d}^l = \mathbf{d}^l - \varrho^i \mathbf{y}^i$;
			\EndFor
			\State $\mathbf{d}^l = \frac{ \langle \mathbf{s}^{l-1},\mathbf{y}^{l-1}\rangle}{ \langle \mathbf{y}^{l-1},\mathbf{y}^{l-1}\rangle} \mathbf{d}^l$;
			\For{$i = 1: 1 : m$}
			\State $\beta = \delta^i \langle \mathbf{y}^i, \mathbf{d}^l \rangle$;		
			\State $\mathbf{d}^l = \mathbf{d}^l + (\varrho^i - \beta)\mathbf{s}^i$;
			\EndFor
			\Ensure $\mathbf{d}^l = -\mathbf{d}^l$
		\end{algorithmic}
	\end{algorithm}
	
	After the direction $\mathbf{d}^l$ is obtained, the variable can be updated as $\mathbf{X}^{l+1} = \mathcal{R}_{\mathbf{X}^l}(\alpha^l\mathbf{d}^l)$ via the line-search strategy. To ensure monotonicity, the line-search update strategy can be implemented as
	\begin{equation}	
		g(\mathbf{X}^{l+1}) \leq g(\mathbf{X}^{l}) +\sigma \gamma^n \tau_l \langle\operatorname{grad}g(\mathbf{X}^{l}), \mathbf{d}^{l} \rangle,
		\label{Amijo}
	\end{equation}
	where $\sigma,\gamma \in (0,1)$, and $\tau_l$ is a relative large initial step size. By increasing $n$ until \eqref{Amijo} is satisfied, a proper step size can be obtained as $\alpha^l = \gamma^n\tau_l$ \cite{rmo}. After the variable is updated, the medium variables $(\mathbf{s}_\mathbf{X}^l,\mathbf{y}_{\mathbf{X}}^l,\delta^l)$ can be calculated. However, to guarantee symmetric positive definiteness of the Hessian approximation, the cautious update can be implemented to decide whether to store the medium variables for the following iterations \cite{rbfgs}. The cautious update condition is given as
	\begin{equation}
		\label{cautious}
		{\langle \mathbf{s}_\mathbf{X}^l,\mathbf{y}_\mathbf{X}^l \rangle} \geq 10^{-4} {\langle \mathbf{s}_\mathbf{X}^l,\mathbf{s}_\mathbf{X}^l \rangle}\lVert\operatorname{grad}g(\mathbf{X}^l)\rVert.
	\end{equation}
	
	For implementing \textbf{Algorithm 1} in the following iteration, the other medium variables $(\mathbf{s}_\mathbf{X}^i,\mathbf{y}_\mathbf{X}^i,\delta^i),\forall i\neq l$ stored in the memory should be transported to the tangent space of the current point $\mathbf{X}^{l+1}$. Since the memory size is limited, the earliest stored medium variable set should be removed if the memory is full. Combining the above, the algorithm for solving \eqref{Qsrs} is summarized in \textbf{Algorithm 2}.
	\begin{algorithm}[t]
		\caption{The RBFGS algorithm for solving \eqref{Qsrs} with fixed $\rho$ and $u$.}
		\begin{algorithmic}[1]
			\Require $\mathbf{X}^0$, empty memory $\mathbf{M}$, convergence threshold $\epsilon$.
			\For{$l=0: \operatorname{Maxiter}$}
			\State Obtain $\mathbf{d}_\mathbf{X}^l$ by \textbf{Algorithm 1};
			\State Obtain $\alpha^l$ and update $\mathbf{X}^{l+1}$ by \eqref{Amijo};
			\If{$\lVert\mathbf{X}^{l+1}-\mathbf{X}^{l}\rVert \leq \epsilon$}
			\State $\operatorname{Break}$;
			\EndIf
			\State Obtain $(\mathbf{s}^l_\mathbf{X},\mathbf{y}^l_\mathbf{X},\delta^l)$ by \eqref{s} and \eqref{y};
			\State Obtain the number of the stored memories $m$;
			\If{\eqref{cautious} is satisfied}
			\If{$m=M_m$}
			\For{$i=1:m-1$}
			\State $\mathbf{s}_\mathbf{X}^i = \mathcal{T}_{\mathbf{X}^{l+1}}\mathbf{s}_\mathbf{X}^{i+1}$, $\mathbf{y}_\mathbf{X}^i = \mathcal{T}_{\mathbf{X}^{l+1}}\mathbf{y}_\mathbf{X}^{i+1}$;
			\State $\mathbf{M}_i = (\mathbf{s}^i_\mathbf{X},\mathbf{y}^i_\mathbf{X},\delta^{i+1})$;
			\EndFor			
			\State Store $\mathbf{M}_{m}=(\mathbf{s}^l_\mathbf{X},\mathbf{y}^l_\mathbf{X},\delta^l)$ in $\mathbf{M}$;
			\Else
			\For{$i=1:m$}
			\State $\mathbf{s}_\mathbf{X}^i = \mathcal{T}_{\mathbf{X}^{l+1}}\mathbf{s}_\mathbf{X}^i$, $\mathbf{y}_\mathbf{X}^i = \mathcal{T}_{\mathbf{X}^{l+1}}\mathbf{y}_\mathbf{X}^i$;
			\State $\mathbf{M}_i = (\mathbf{s}^i_\mathbf{X},\mathbf{y}^i_\mathbf{X},\delta^i)$;
			\EndFor		
			\State Store $\mathbf{M}_{m+1}=(\mathbf{s}^l_\mathbf{X},\mathbf{y}^l_\mathbf{X},\delta^l)$ in $\mathbf{M}$;
			\EndIf
			\Else
			\For{$i=1:m$}
			\State $\mathbf{s}_\mathbf{X}^i = \mathcal{T}_{\mathbf{X}^{l+1}}\mathbf{s}_\mathbf{X}^i$, $\mathbf{y}_\mathbf{X}^i = \mathcal{T}_{\mathbf{X}^{l+1}}\mathbf{y}_\mathbf{X}^i$;
			\EndFor
			\EndIf
			\State $l=l+1$;
			\EndFor
			\Ensure $\mathbf{X} = \mathbf{X}^{l+1}$.
		\end{algorithmic}
	\end{algorithm}
	
	\subsection{REP Method for Finding Feasible Solution}
	The values of $\rho$ and $u$ in \eqref{Qsrs} are crucial for obtaining the feasible solution of \eqref{Qsr3} via solving \eqref{Qsrs}. The inequality conditions are likely to hold if $\rho$ is sufficiently large and $u$ is small enough. However, too large $\rho$ leads to slow convergence and numerical difficulties, while the obtained solution of $\eqref{Qsrs}$ can be far from feasible if $\rho$ is too small. Instead, a practical approach is the REP method, where proper values of $\rho$ and $u$ are found by solving \eqref{Qsrs} and increasing and decreasing initial values iteratively \cite{rep}. Specifically, after obtaining the solution of \eqref{Qsrs}, we increase $\rho$ as $\rho=\theta_\rho \rho$ with $\theta_\rho >1$ if \eqref{ineq} is not satisfied. The smoothing parameter is decreased iteratively as $u=\max\{u_{min},\theta_u u\}$ with $\theta_u \in (0,1)$ to increase the precision of LSE, where $u_{min}$ is the lower bound. Thus, a feasible solution of \eqref{Qsr3} can be found by iteratively solving \eqref{Qsrs} via \textbf{Algorithm 2} with the fixed $\rho$ and $u$ and updating $\rho$ and $u$ until \eqref{ineq} is satisfied. To improve the accuracy of the solution, the convergence threshold of \textbf{Algorithm 2} can be decreased in iterations as $\epsilon = \max\{\epsilon_{min},\theta_\epsilon \epsilon\}$ with $\theta_\epsilon\in(0,1)$, where $\epsilon_{min}$ is the lower bound. Combining the REP method and the RBFGS algorithm, the PRMO for solving \eqref{Qsr3} is summarized as \textbf{Algorithm 3}.
	\begin{algorithm}[t]
		\caption{PRMO method for the sum-rate maximization exploiting REP}
		\begin{algorithmic}[1]
			\Require Initial point $\mathbf{X}^1$, $\rho^1$, $u^1$, $\epsilon^1$, $\theta_\rho>1$, $\theta_u\in(0,1)$, $\theta_\epsilon\in(0,1)$, lower bound $u_{min}, \epsilon_{min}$, convergence threshold $\tau$.
			\Repeat
			\State \parbox[t]{\dimexpr\linewidth-\algorithmicindent}{Obtain a solution $\mathbf{X}^{l+1}$ by solving \eqref{Qsrs} via \textbf{Algorithm 2} with warm-start at $\mathbf{X}^l$, fixed $u^l$, $\rho^l$, and convergence threshold $\epsilon^l$;} 
			\If{$h_i(\mathbf{X}^{l+1})>0, \exists i \in \mathcal{I}$}
			\State $\rho^{l+1}=\theta_\rho \rho^l$;
			\State $\mathbf{X}^{l+1}=\mathbf{X}^l$;
			\Else
			\State $\rho^{l+1}=\rho^l$;
			\EndIf
			\State $u^{l+1} = \theta_u u^l$, $\epsilon^{l+1} = \theta_\epsilon \epsilon^l$;			
			\State $l=l+1$;
			\Until{$\lVert \mathbf{X}^{l+1}-\mathbf{X}^l\rVert < \tau$ and $h_i(\mathbf{X}^{l+1})\leq0, \forall i \in \mathcal{I}$ and $u^{l+1}\leq u_{min}$ and $\epsilon^l\leq \epsilon_{min}$.}
			\Ensure $\mathbf{X}^{l+1}=[\mathbf{W}^{l+1},\mathbf{o}^{l+1},\mathbf{p}^{l+1}]$, $\mathbf{W} = \mathbf{W}^{l+1}$, $\mathbf{t}=p_B(\mathbf{o}^{l+1})$, $\mathbf{u}=p_I(\mathbf{p}^{l+1})$.
		\end{algorithmic}
	\end{algorithm}

	\subsection{Computational Complexity Analysis}
	The complexity of \textbf{Algorithm 2} mainly lies in the calculation of the Euclidean gradients and the P-RBFGS direction. Observing the matrix multiplication with large dimensions, the computational complexity for calculating $\nabla_{\mathbf{W}^*}g(\mathbf{X})$ and $\nabla_{\boldsymbol{\phi}^*}g(\mathbf{X})$ are estimated as $\mathcal{O}(K^3M)$ and $\mathcal{O}(K^3N^2M)$, respectively. For calculating $\nabla_{\mathbf{o}}g(\mathbf{X})$ and $\nabla_{\mathbf{p}}g(\mathbf{X})$, the complexity orders are estimated as $\mathcal{O}\bigl(K(LMN+M^2N+K(MN+N^2))\bigr)$ and $\mathcal{O}\bigl(K(LMN+MN^2+K(N^2M))\bigr)$, respectively. It can be observed that the complexity of \textbf{Algorithm 1} is $\mathcal{O}\bigl(M_m(MK+N)\bigr)$. Thus, the computational complexity of \textbf{Algorithm 3} is estimated as $\mathcal{O}\bigl(I\bigl(M_m(MK+N)+K^3N^2M+M^2KN+KLMN\bigr)\bigr)$ per iteration, where $I$ is the iteration number of \textbf{Algorithm 2}.

	\subsection{Convergence Analysis}
	The line-search strategy \eqref{Amijo} guarantees the monotone of \textbf{Algorithm 2} \cite{conv}, that is if $\mathbf{X}^{l}$ is bounded over the PRMS $\mathcal{M}$, there exists $c > 0$  such that
	\begin{equation}	
		g(\mathbf{X}^{l+1}) - g(\mathbf{X}^{l}) \leq   c \langle\operatorname{grad}g(\mathbf{X}^{l}), \mathbf{d}_\mathbf{X}^{l} \rangle,
	\end{equation}
	then we can conclude that $g(\mathbf{X}^{l+1}) \leq g(\mathbf{X}^{l})$ and $g(\mathbf{X})$ is non-increasing (and so is $g'(\mathbf{X}')$). The obtained rate $r_k, \forall k \in \mathcal{K}$ is upper bounded due to the power constraint \eqref{Cp}. Thus, with fixed $\rho$ and $u$, \textbf{Algorithm 2} is guaranteed to converge in each iteration of \textbf{Algorithm 3}. 
	
	To prove the convergence of \textbf{Algorithm 3}, we first give the definition of linear independence constraint qualifications (LICQ) conditions. As given in \cite{licq}, LICQ conditions are satisfied at $\mathbf{X}$ if $\{\operatorname{grad}h_i(\mathbf{X}),i\in \mathcal{A}(\mathbf{Q}) \cap \mathcal{N}\}$ are linearly independent in $\mathrm{T}_\mathbf{X}\mathcal{M}$, where $\mathcal{A}(\mathbf{X})$ denotes the active set of constraints, that is $\mathcal{A}(\mathbf{X}) = \{i\in\mathcal{I}|h_i(\mathbf{X}) = 0\}$. Assume that a large enough $\bar{\rho}$ is found and fixed. Theoretically, if the stopping criterion parameters are set as $\tau = d_{min} = u_{min} = 0$, it can be concluded that if the sequence $\{\mathbf{X}^l\}$ produced by \textbf{Algorithm 2} admits a feasible limit point $\bar{\mathbf{X}}$ that satisfies the LICQ condition, then $\bar{\mathbf{X}}$ satisfies the KKT conditions for \eqref{Qsrs}. The proof is given in Appendix B.

	\section{Simulation Results and Discussion}
	In this section, we provide simulation results to prove the effectiveness of the proposed MA-IRS-aided MU-MISO communication system and the PRMO method for the joint beamforming and antenna position optimization. Unless otherwise noted, the simulation parameters are as follows. Consider the Cartesian coordinate with the unit of 1m. The BS and IRS are located at $(0,0,0)$ and $(10,0,30)$, respectively. The number of UE is set as $K=3$. The UEs are randomly located in a square region centered at $(0,-10,30)$ with an edge length of 20 in the $x$-$z$ plane. The carrier frequency is set as 5 GHz, then the free space path loss of the BS-IRS and IRS-UE channels is given by $L(d) = -46-20 \log (d)$ dB, where $d$ is channel distance \cite{los}. The channel responses are generated by $\sigma_{G,l}\sim\mathcal{CN}(0,\mu_G/L)$ and $\sigma_{k,l}\sim\mathcal{CN}(0,\mu_k/L)$, where $\mu_G$ and $\mu_k$ are the path losses of the BS-IRS channel and the IRS-UE channel for the $k$-th UE, respectively. The noise power is set as $\sigma_n^2=-120$ dBm. The path angles $\phi_{t,l}$ and $\theta_{r,l},\phi_{r,l},\forall l$ are randomly generated from uniform distribution over $[0,\pi]$. 	
	
	For the initialization, the MAPs of the BS are initialized as that of FPA with half wavelength antenna distance. The MAPs of the IRS are initialized via the circle packing scheme to guarantee the initial MAPs are sufficiently separated \cite{MA9}. For the proposed MA-IRS, the phase shifts are fixed as $\phi_i = 2\pi, \forall i \in \mathcal{N}$. The precoding matrix $\mathbf{W}$ is initialized via the ZF beamformer to obtain a fair initial power allocation \cite{MA7,MA10}, which is given by $\mathbf{W}_{ini}=\sqrt{P_t/\operatorname{Tr}((\mathbf{H}^H\mathbf{H})^{-1})}\mathbf{H}(\mathbf{H}^H\mathbf{H})^{-1}$, where $\mathbf{H}=[\mathbf{h}_1,\dots,\mathbf{h}_K]$. 
	
	\subsection{Convergence Performance}
	\begin{figure}[t]
		\centering{\includegraphics[width=0.8\columnwidth]{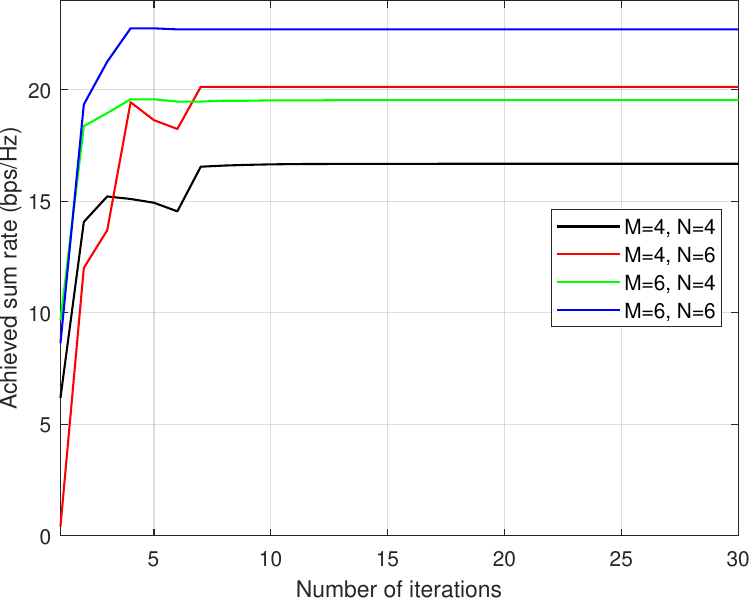}}
		\caption{Convergence performance of the proposed PRMO method with $P_t=30$ dBm, $L=6$, $A_B=4$, $A_I=6$, and $\Gamma = 1$ bps/Hz.}
	\end{figure}
	The convergence performance of the proposed PRMO method over several random channel realizations is presented in Fig. 2. We can observe that the REP for searching proper penalty weight causes fluctuation in sum-rate in the first few iterations, while the sum-rate achieves convergence in about 10 iterations once the feasible penalty weight is found for satisfying the constraints. In addition, it is demonstrated that a larger number of MA leads to a higher achieved sum-rate. Since the MA-IRS can control both BS-IRS and IRS-UE channels, increasing the number of MA in the IRS yields greater performance gains than increasing the number of MA in the BS.
	
	\subsection{Performance Comparisons and Analysis}
	In this section, we provide the performance comparison to evaluate the effectiveness of the MA-IRS-aided MU-MISO communication system and the proposed PRMO method (denoted as ``\textbf{Proposed}"). Specifically, we consider two modes for the proposed MA-IRS. For the first mode, the positions and phase shifts of the movable elements are jointly optimized as introduced previously (denoted as ``\textbf{OPS}"). For the second mode, only the positions of the movable elements are optimized, while the phase shifts are fixed as $2\pi$ to reduce computational overhead (denoted as ``\textbf{FPS}"). The IRS phase shifts are assumed to be continuous within the schemes that optimize $\boldsymbol{\phi}$ (denoted as ``\textbf{CPS}"). Considering the practical implementation of the IRS, we include the results obtained with discrete IRS phase shifts (denoted as ``\textbf{DPS}"). Besides, we denote the random phase shifts without optimization as ``\textbf{RPS}". We assume that for DPS, each reflecting element only takes discrete phase shifts from $\kappa$ values, which is denoted as
	\begin{equation}
		\phi'_n \in \{0,\frac{2\pi}{\kappa},\dots,\frac{2\pi(\kappa-1)}{\kappa}\}, \forall n \in\mathcal{N}.
	\end{equation}
	Once the CPS $\phi_n$ is obtained, the DPS is determined by its projection into the DPS set. In the simulation, we set the number of the DPS as $\kappa = 16$ for each IRS reflecting element. In the comparisons, we consider the following baseline schemes that have been investigated in existing works.
	\begin{itemize}
		\item \textbf{FPA}: The IRS-aided MU-MISO communication where BS and IRS are equipped with FPA with fixed initial MAP \cite{IRS4}. The sum-rate maximization is tackled by the joint active and passive beamforming of precoding matrix $\mathbf{W}$ and IRS phase shift $\boldsymbol{\phi}$.
		\item \textbf{FPA-MA}: The MA-IRS-aided MU-MISO communication where the BS is equipped with FPA \cite{MAIRS2}. The problem is tackled by the joint beamforming of $\mathbf{W}$ and the MAP optimization of $\mathbf{u}$.
		\item \textbf{MA-FPA}: The FPA-IRS-aided MU-MISO communication where the BS is equipped with MA \cite{MAIRS1}. The sum-rate maximization is solved by the joint beamforming of $\mathbf{W}$ and $\boldsymbol{\phi}$ and MAP optimization of $\mathbf{t}$.	
	\end{itemize}
	
	It should be note that the methods in the existing works cannot be used directly to solve the beamforming and the joint beamforming and antenna position optimization problems for the schemes due to the different problem formulations. Nevertheless, the problems can be tackled via the proposed PRMO method by modifying the construction of the PRMS. According to the analysis in Section III-E, the computational complexities of the schemes are compared in Table I. It can be observed that with fixed phase shifts, the computation overhead can be reduced for the schemes. 
	\begin{table}[t]
		\caption{Computational Complexity Comparison of the Schemes}
		\centering
		\begin{tabular}{|c|c|} \hline
			Scheme & Complexity order (per iteration)  \\ \hline
			Proposed-OPS& \makecell[c]{$\mathcal{O}\bigl(I(M_m(MK+N)+K^3N^2M$ \\ $+M^2KN+KLMN)\bigr)$} \\ \hline
			Proposed-FPS  &	\makecell[c]{$\mathcal{O}\bigl(I(M_m(MK+N)+K^3M+N^2K^2M$ \\ $+M^2KN+KLMN)\bigr)$}  \\ \hline
			FPA-MA-OPS  & \makecell[c]{$\mathcal{O}\bigl(I(M_m(MK+N)+K^3N^2M$ \\ $+KLMN)\bigr)$} \\ \hline
			FPA-MA-FPS  & \makecell[c]{$\mathcal{O}\bigl(I(M_m(MK+N)+K^3M+N^2K^2M$ \\ $+KLMN)\bigr)$} \\ \hline
			MA-FPA  & \makecell[c]{$\mathcal{O}\bigl(I(M_m(MK+N)+K^3N^2M$ \\ $+M^2KN+KLMN)\bigr)$} \\ \hline
			FPA  & $\mathcal{O}\bigl(I(M_m(MK+N)+K^3N^2M)\bigr)$ \\ \hline
		\end{tabular}
	\end{table}

	The sum-rate obtained by the proposed and baseline schemes versus the transmit power $P_t$ is presented in Fig. 3. We can observe that compared to RPS, the conventional FPA-IRS can enhance the sum-rate by optimizing the phase shifts of the IRS. In contrast, employing MA further improves the achieved sum-rates. Note that for the proposed scheme which applies MA at both the BS and IRS achieves the best performance. For MA-IRS, compared with the FPS mode, the OPS mode provides only a marginal performance gain. However, when employing DPS, the imprecise phase control leads to performance degradation, with the loss increasing as $P_t$ increases. Similar performance degradation is also observed for FPA and MA-FPA schemes. In comparison, the proposed scheme with FPS mode achieves performance comparable to that of OPS mode with precise phase control, while outperforming that with imprecise phase control and other schemes. This demonstrates that the proposed MA-IRS can achieve enhanced performance solely by adjusting the position of each element, thereby reducing the hardware overhead associated with precise phase shift control.
	\begin{figure}[t]
		\centering{\includegraphics[width=0.8\columnwidth]{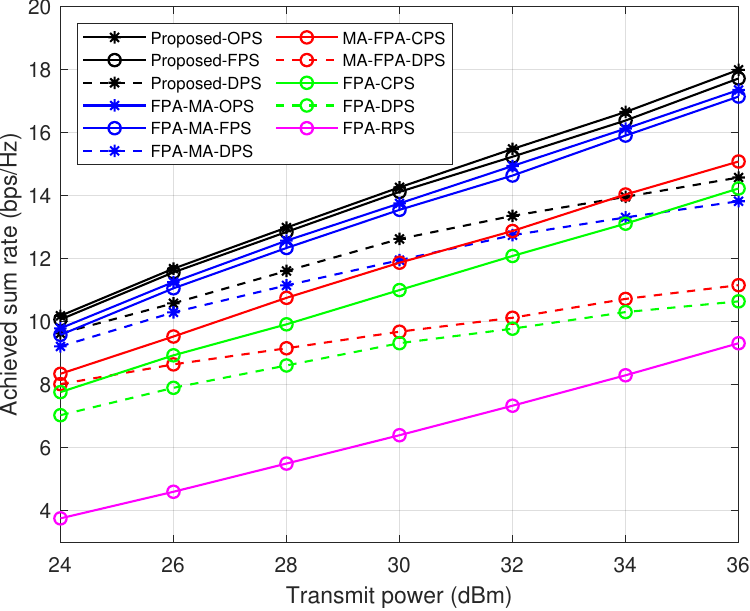}}
		\caption{Sum-rate versus transmit power $P_t$ with $M=4$, $N=8$, $L=8$, $A_B/\lambda=4$, $A_I/\lambda=6$, and $\Gamma = 1$ bps/Hz.}
	\end{figure}
	\begin{figure}[t]
		\centering{\includegraphics[width=0.8\columnwidth]{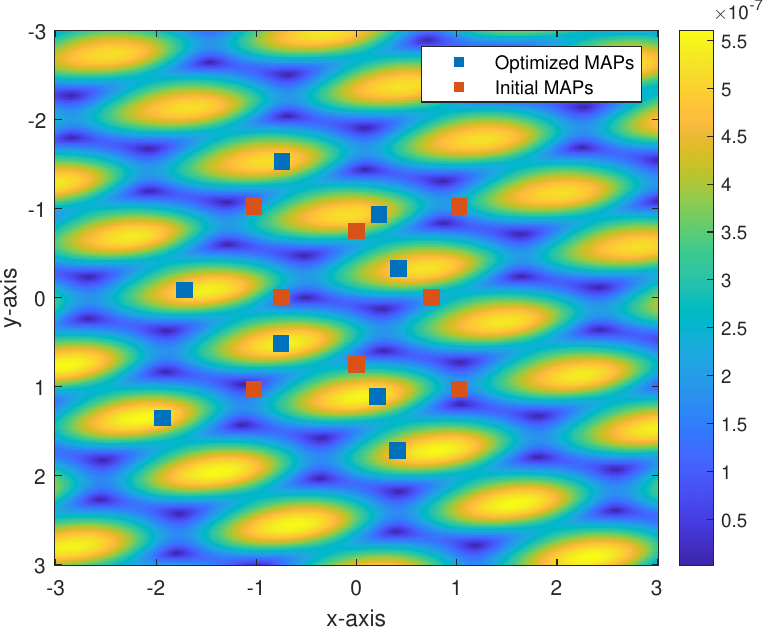}}
		\caption{Channel power gain for one UE with respect to the MAPs of the MA-IRS.}
	\end{figure}
	The proposed scheme outperforms the other schemes because both the BS-IRS and IRS-UE channels can be adjusted by changing the positions of the IRS reflecting elements, thereby enhancing the channel gain. For the $k$-th UE, the channel power gain is given as $\lVert\mathbf{h}_k\rVert^2=\boldsymbol{\phi}^H\operatorname{diag}(\mathbf{f}^H_{k})\mathbf{G}\mathbf{G}^H\operatorname{diag}(\mathbf{f}_{k})\boldsymbol{\phi}$,
	where $\boldsymbol{\phi} = [e^{j\phi_1},\dots, e^{j\phi_N}]^H$ and $\phi_n\in[0,2\pi),\forall n \in \mathcal{N}$. If $\phi_n$ is uniformly distributed over $[0,2\pi)$, the expected value of the channel power gain is given by $\mathbb{E}(\lVert\mathbf{h}_k\rVert^2)=\operatorname{Tr}(\mathbf{G}^H\operatorname{diag}(\mathbf{f}_{k})\operatorname{diag}(\mathbf{f}^H_{k})\mathbf{G})$. According to the Rayleigh-Ritz theorem, the theoretical maximum value of the channel power gain is $N\lambda_{\max}$, where $\lambda_{\max}$ is the maximum eigenvalue of matrix $\operatorname{diag}(\mathbf{f}^H_{k})\mathbf{G}\mathbf{G}^H\operatorname{diag}(\mathbf{f}_{k})$. For the FPA-IRS-aided communications, although the phase shift could be optimized to enhance the channel power gain, the expected and maximum channel power gains cannot be further improved since the channel matrices including $\mathbf{G}$ and $\mathbf{f}_k,\forall k \in \mathcal{K}$ are fixed. However, by optimizing the antenna positions of BS and the proposed MA-IRS, the values of the elements within $\mathbf{G}$ and $\mathbf{f}_k,\forall k \in \mathcal{K}$ are adjustable and can be optimized to enhance the exception and maximum value of the channel power gain. Besides, since the value of each channel element can be adjusted by controlling the positions of the MA, the signals arriving at the UE through each reflected path can achieve higher coherence, which to some extent can compensate for the function of phase shift optimization.
	
	To illustrate the effect of the movable elements of the MA-IRS on enhancing the channel power gain, for the proposed scheme with FPS mode, the channel power gain with respect to the MAPs of the MA-IRS for one UE is shown in Fig. 4. We can find that each element of the MA-IRS can move to a near position with higher channel gain via the proposed PRMO, thus enhancing the system performance without controlling the phase shifts. 
	
	\begin{figure}[t]
		\centering
		{\includegraphics[width=0.8\columnwidth]{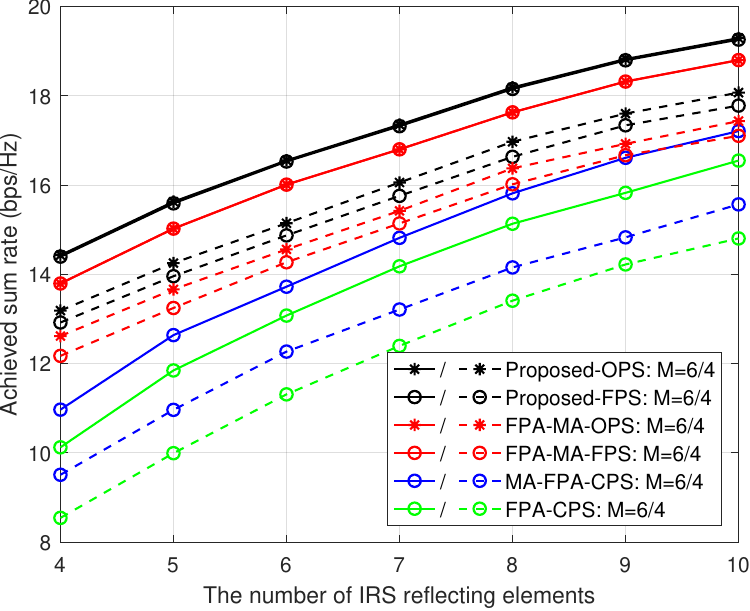}}
		\caption{Sum-rate versus the number of IRS elements with $P_t=30$ dBm, $L=6$, $A_B/\lambda = 4$, $A_I/\lambda=6$, $\Gamma=1$ bps/Hz.}
	\end{figure}
	\begin{figure}[t]
		\centering{\includegraphics[width=0.8\columnwidth]{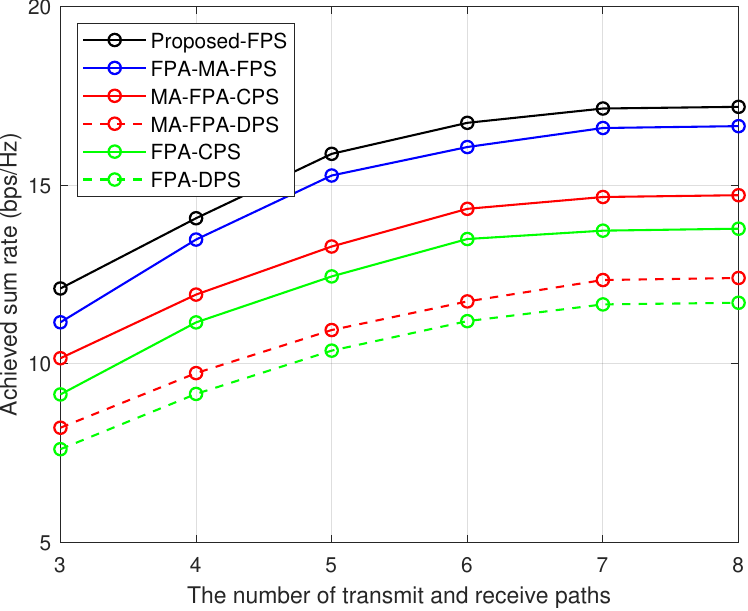}}
		\caption{Sum-rate versus channel path number $L$ with $P_t = 30$ dBm, $M=4$, $N=8$, $A_B/\lambda=4$, $A_I/\lambda=6$, and $\Gamma = 1$ bps/Hz.}
	\end{figure}
	As observed, many positions with high channel gain may exist within the MA regions, which can be further explored by more MAs. The impact of the numbers of MA on system performance is shown in Fig. 5. The results show that increasing the number of antennas enhances the achieved sum-rate for all the schemes, while the proposed scheme achieves the highest sum-rate. With a larger number of MAs, the MA-IRS can move more elements to the positions with high channel gain, thus achieving higher performance. Similarly, we can observe that the proposed scheme achieves only marginal performance gains in the OPS mode compared to the FPS mode. Moreover, as the number of MAs increases, more channel paths with high power gain further diminishes the performance gains obtained through phase shift optimization. Typically, when $M=4$ and $N=8$, the proposed scheme with FPS mode achieves $4.6\%$, $19.1\%$, and $29.4\%$ higher sum-rate compared with the ``FPA-MA" with FPS, ``MA-FPA" with CPS, and the ``FPA" with CPS, respectively.

	\begin{figure}[t]
		\centering{\includegraphics[width=0.8\columnwidth]{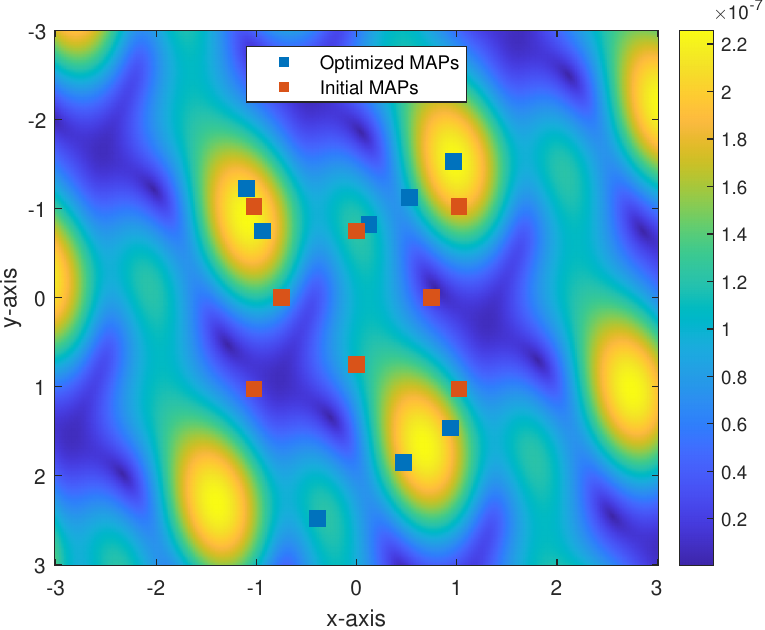}}
		\caption{Channel power gain for one UE with respect to the MAPs of the MA-IRS with $L=3$.}
	\end{figure}
	
	The BS-IRS and IRS-UE channels are also characterized by the number of transmit and receive channel paths $L$. Fig. 6 illustrates the impact of  $L$ on system performance. We can find that the proposed scheme achieves the highest sum-rate with different $L$. Besides, as the number of channel paths increases, the sum-rate obtained by all the schemes increases. This is because a larger number of local maxima of channel gain is introduced by the more pronounced small-scale fading in the spatial domain with increased number of channel paths \cite{MA3,MA9}. The channel gain with $L=3$ for one UE within the MA-IRS region is shown in Fig. 7. Compared with Fig. 6, it can be observed that fewer local maxima of channel gain can be exploited. Due to the minimum MA distance constraint, not all the MAs can move to positions with high channel gain, leading to performance loss.
	
	\begin{figure}[t]
		\centering{\includegraphics[width=0.8\columnwidth]{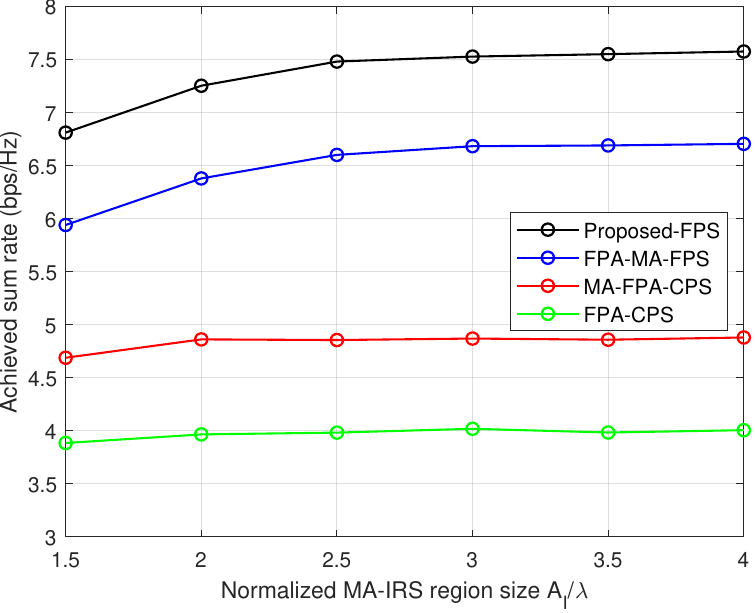}}
		\caption{Sum-rate versus normalized region size $A_{I}/\lambda$ with $P_t = 24$ dBm, $M=4$, $N=8$, $L=8$, $A_B/\lambda=3$, and $\Gamma = 1$ bps/Hz.}
	\end{figure}
	\begin{figure}[t]
		\centering
		\subfigure[$A_I/\lambda=1.5$]{
			\begin{minipage}[t]{0.8\linewidth}
				\centering
				\includegraphics[width=1\textwidth]{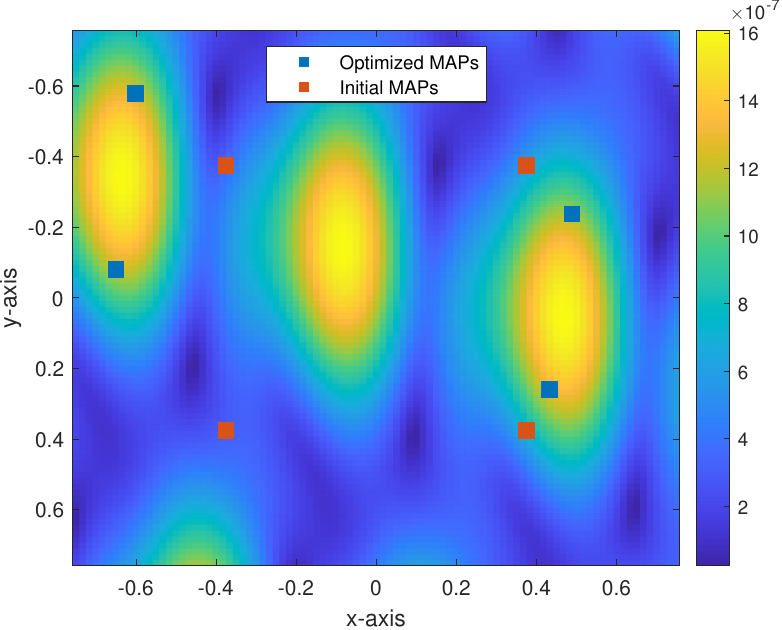}
		\end{minipage}}\\
		\subfigure[$A_I/\lambda=3$]{
			\begin{minipage}[t]{0.8\linewidth}
				\centering
				\includegraphics[width=1\textwidth]{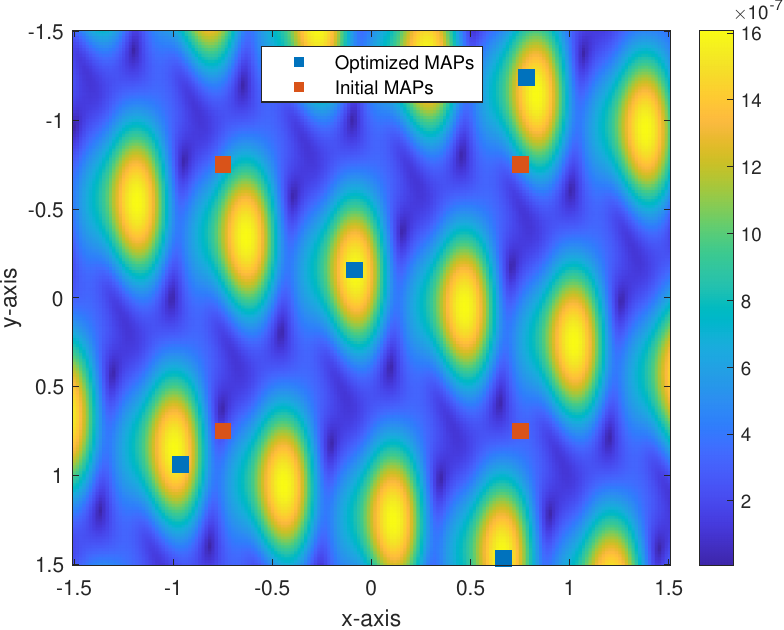}
		\end{minipage}}
		\caption{Channel power gain for one UE with respect to the MAPs of the MA-IRS with different region sizes.}
	\end{figure}	
	\begin{figure}[t]
		\centering{\includegraphics[width=0.8\columnwidth]{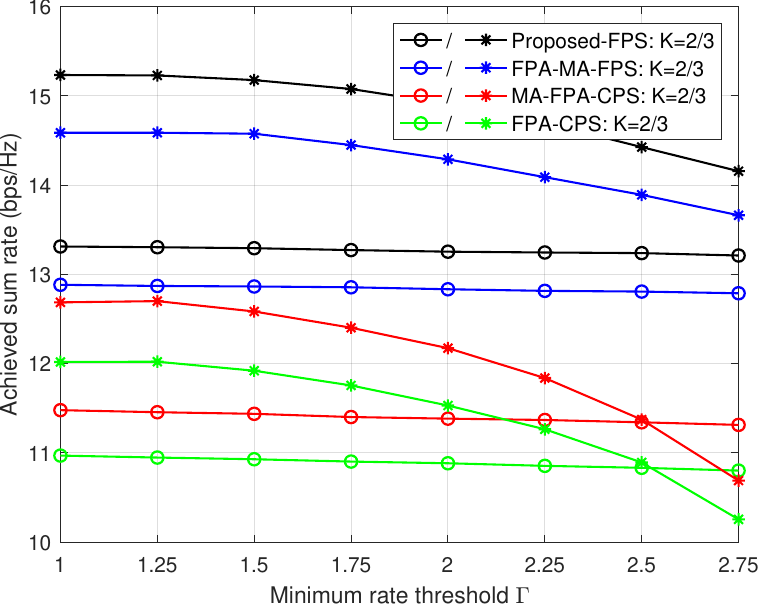}}
		\caption{Sum-rate versus minimum rate threshold $\Gamma$ with $P_t = 28$ dBm, $M=4$, $N=8$, $L=6$, $A_B/\lambda=4$,  $A_I/\lambda=6$.}
	\end{figure}
	The lack of channel gain maxima can be tackled by increasing the region size of the MA-IRS. The achieved sum-rate versus the normalized MA region size of IRS $A_I/\lambda$ for the proposed MA-IRS-aided system and the baseline schemes is demonstrated in Fig. 8. It is shown that the MA-IRS outperforms FPA-IRS in sum-rate maximization, and the optimized sum-rate increases with larger region size of the MA-IRS. As shown in Fig. 9(a), with a relative small region size, not all the elements can move to the position with high channel gain due to the lack of channel gain maxima and the minimum distance constraint. With a larger region size, it is shown in Fig. 9(b) that all the elements can move to the positions with high channel gain while satisfying the minimum distance. Additionally, the sum-rates achieved by the MA-IRS converge with increasing region size, suggesting that there are sufficient positions with high channel gain within the region, and a maximum sum-rate can be attained with a finite MA region size.

	The impact of the minimum rate threshold $\Gamma$ on the proposed PRMO method with different number of UEs is shown in Fig. 10. It is illustrated that higher number of UEs can lead to higher sum-rate when the minimum rate threshold is low. However, with a given transmit power, the sum-rates obtained by all the schemes decrease with a larger minimum rate threshold $\Gamma$, and the degradation is more pronounced with higher number of UEs. It is because with limited maximum transmit power, for higher number of UEs, more power is allocated to the UEs with poor channel conditions to meet the minimum rate threshold, thus reducing the rate obtained by the UEs with good channel conditions and lowering the optimized sum-rate. This indicates that in practical scenarios, the maximum power should be determined jointly based on channel conditions, the number of UEs, and the minimum rate requirements. Compared with FPA-IRS, the proposed MA-IRS further enhances the channel conditions for the UEs, thus achieving higher sum-rates with different minimum rate thresholds. 
	\begin{figure}[t]
		\centering{\includegraphics[width=0.8\columnwidth]{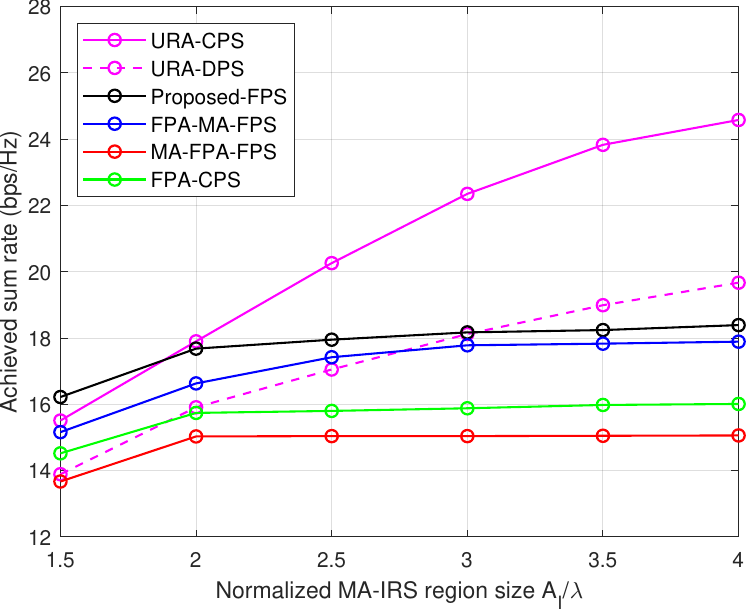}}
		\caption{Sum-rate versus normalized region size $A_I/\lambda$ with $P_t = 32$ dBm, $M=4$, $N=10$, $L=6$, $A_B/\lambda=3$, and $\Gamma=1$ bps/Hz.}
	\end{figure}
	\begin{table}[t]
		\caption{Comparison of Execution Time (s)}
		\centering
		\begin{tabular}{|c|c|c|c|c|} \hline
			\diagbox{Scheme}{$A_I/\lambda$} & 1.5 & 2 & 3 & 4   \\ \hline
			Proposed ($N=10$)&	8.53 & 8.23 & 7.65 & 7.55\\ \hline
			URA& 8.82 & 12.68 & 20.83 & 43.97  \\ \hline
		\end{tabular}
	\end{table}
	
	Conventional FPA-IRS could employ large number of reflecting elements to achieve higher system performance. Although the region size of the IRS considered in this study is limited, for conventional FPA-IRS, the reflecting elements can be arranged more densely to deploy a larger number of elements, thereby achieving higher system performance. Meanwhile, it is shown in Table I that even for FPA-IRS, the computational complexity increases with the number of elements, resulting in greater computational overhead. In Fig. 11, we compare the performance of the proposed MA-IRS with $N=10$ to that of the FPA-IRS employing a uniform rectangular array under the same region size, which is denoted as ``URA". In this setup, the elements are arranged with a half-wavelength spacing, leading to $(2A_I/\lambda + 1)^2$ reflecting elements, which is substantially higher than that for the proposed MA-IRS. Considering the difficulty of phase accuracy for a large number of IRS elements, we also employ DPS to quantize the phase shifts. Table 2 compares the execution time required for convergence between the proposed scheme and the URA scheme under different region sizes. It can be observed that when the region size is limited, the proposed MA-IRS can achieve a higher or comparable sum rate with fewer elements and fixed phase compared to the URA scheme. As the region size increases, the number of IRS elements grows exponentially for the URA scheme, enabling it to achieve higher performance. However, optimizing a large number of phase shifts significantly increases the execution time, and substantial performance degradation is observed when DPS is employed. In contrast, the proposed MA-IRS requires less execution time since the position with high channel gain could be found more easily. The result indicates that when the IRS region size is constrained, the MA-IRS could serve as a more advantageous scheme.

	\subsection{Impact of Imperfect FRI}
	We assume that the FRI including the AoDs, AoAs, and the FRMs of the channel paths are perfectly known for the above simulation results. However, in practical communication systems, perfect FRI is challenging to obtain due to limited CE overhead. In this section, we evaluate the impact of imperfect FRI on the performance of the proposed MA-IRS-aided communications and the PRMO method. First, we consider the errors in the angles including the AoDs and AoAs of the channel paths. Specifically, as depicted in Section II-A, to obtain the channel information, channel information acquisition requires the AoD $\phi_{t,l}$ and the AoA of $\theta_{r,l}$ and $\phi_{r,l}$, for all $l=1,\dots,L$. We assume that the estimation errors between the true and estimated angles are independent random variables following uniform distribution with maximum angle error $\mu$, which can be denoted as $\hat{\phi}_{t,l}-\phi_{t,l}\sim\mathcal{U}[-\mu/2,\mu/2]$, $\hat{\theta}_{r,l}-\theta_{r,l}\sim\mathcal{U}[-\mu/2,\mu/2]$, and $\hat{\phi}_{r,l}-\phi_{r,l}\sim\mathcal{U}[-\mu/2,\mu/2]$, where $\hat{\phi}_{t,l}$, $\hat{\theta}_{r,l}$, and $\hat{\phi}_{r,l}$ denote the estimated angles, and $\mathcal{U}$ denotes the uniform distribution. Additionally, the channel is characterized by the FRMs $\mathbf{\Sigma}_\mathrm{G} =\operatorname{diag}([\sigma_{G,1},\dots,\sigma_{G,L}]^T)$ and $\mathbf{\Sigma}_{\mathrm{f},k} =\operatorname{diag}([\sigma_{k,1},\dots,\sigma_{k,L}]^T), \forall k \in \mathcal{K}$ with complex responses $\sigma_{G,l}$ and $\sigma_{k,l}$. For all $l=1,\dots,L$, we denote the normalized estimation error of the FRMs as $\frac{\hat{\sigma}_{G,l}-\sigma_{G,l}}{\lvert \hat{\sigma}_{G,l} \rvert}\sim\mathcal{CN}(0,\nu)$ and $\frac{\hat{\sigma}_{k,l}-\sigma_{k,l}}{\lvert \hat{\sigma}_{k,l} \rvert}\sim\mathcal{CN}(0,\nu),\forall k\in\mathcal{K}$, where $\hat{\sigma}_{G,l}$ and $\hat{\sigma}_{k,l}$ denote the estimated complex responses, and $\nu$ is the maximum variance of the normalized FRM error.

	We first implement the PRMO based on the estimated angles with perfect FRMs, and the impact of the maximum angle error on the real sum-rate is shown in Fig. 11. We can observe that the sum-rate decreases with larger angle errors due to the misalignment between the angles of the estimated and true channel paths. With different value of $\mu$, the MA-IRS-aided systems achieves higher sum-rate. Typically, with the MA-enabled BS, the communication aided by the proposed MA-IRS with fixed phase shift and angle error of $\mu=0.04$ achieves approximately equal sum-rate compared with that obtained by the conventional FPA-IRS with perfect angle information and optimized phase shifts. The impact of the maximum FRM error on the proposed MA-IRS and PRMO is demonstrated in Fig. 12, where the angle information is assumed to be perfect. We can find that the larger errors in both $\mathbf{\Sigma}_\mathrm{G}$ and $\mathbf{\Sigma}_{\mathrm{f},k}, \forall k \in \mathcal{K}$ lead to decreased optimized sum-rate, since the inaccurate channel information cause by the FRM error. However, the proposed MA-IRS still shows superiority over the conventional FPA-IRS, even with fixed phase shifts. 
	
	\begin{figure}[t]
		\centering{\includegraphics[width=0.8\columnwidth]{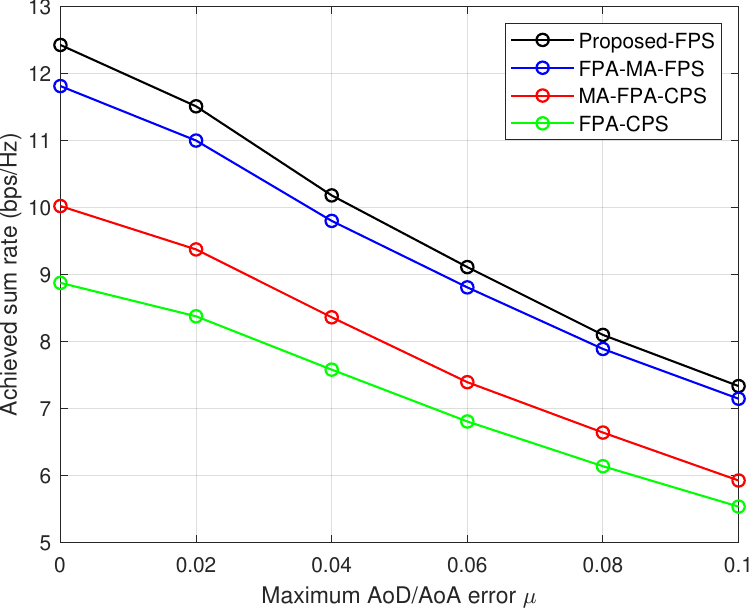}}
		\caption{Sum-rate versus maximum angle error $\mu$ with $P_t = 30$ dBm, $M=4$, $N=6$, $L=4$, $A_B/\lambda=4$, and $A_I/\lambda=4$.}
	\end{figure}
	\begin{figure}[t]
		\centering{\includegraphics[width=0.8\columnwidth]{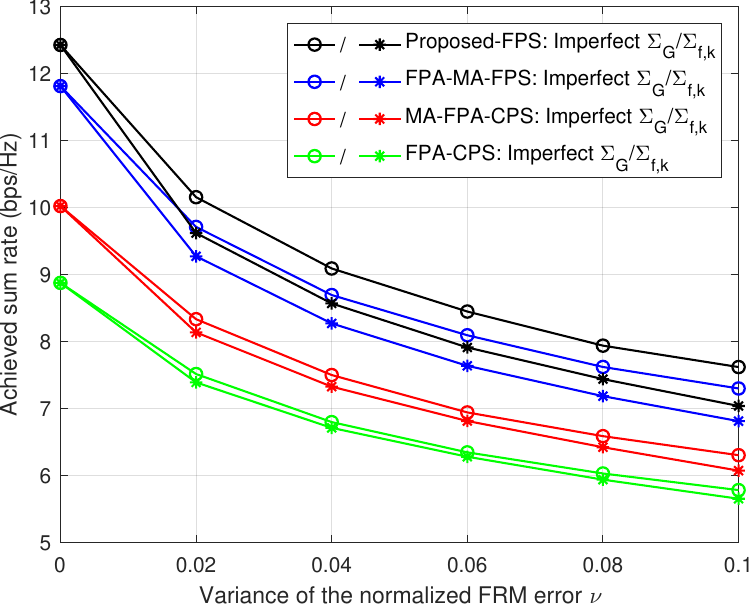}}
		\caption{Sum-rate versus maximum FRM error $\mu$ with $P_t = 30$ dBm, $M=4$, $N=6$, $L=4$, $A_B/\lambda=4$, and $A_I/\lambda=4$.}
	\end{figure}

	\section{Conclusion}
	In this paper, we have proposed a novel MA-IRS-aided MU-MISO communication system and a PRMO method for the joint beamforming and MAP optimization of the system. Be different from the conventional FPA-IRS, the proposed MA-IRS enhances the wireless channel by controlling the positions of the reflecting elements within a given region size. To tackle a highly non-convex minimum-rate constrained sum-rate maximization problem, we reformulated the problem over a constructed PRMS, and the REP method was utilized to tackle the minimum rate and minimum MA distance constraints. Exploiting the derived Riemannian gradients and the RBFGS algorithm, a feasible solution could be obtained via the proposed PRMO method. The simulation results have verified the effectiveness of the proposed MA-IRS-aided system and the PRMO method, and it was demonstrated that the proposed MA-IRS further improves the channel power gain for the UEs, thus outperforming the conventional FPA-IRS in terms of sum-rate maximization. More advanced algorithm could be investigated for the MA-IRS to reduce complexity and improve system performance in the future. Besides, since precisely control of the positions of all movable elements can be challenging with larger element number, future work could focus on MA-IRS with partially movable reflecting elements or movable array that contains multiple elements for the MA-IRS.
	
	\appendices
	\section{Detailed Partial Derivatives}
	Exploiting basic matrix operations \cite{CM}, we can obtain the following partial derivative with respective to $\mathbf{w}_k,\forall k\in\mathcal{K}$:
	\begin{flalign}	
		\frac{\partial r_k}{\partial \mathbf{w}^*_k} = \frac{\partial r_k}{\partial \gamma_k}\frac{\partial \gamma_k}{\partial \mathbf{w}^*_k} = \frac{2}{1+\gamma_k}\cdot\frac{\mathbf{h}_k^H\mathbf{w}_k\mathbf{h}_k}{{\sum_{j\neq k}\lvert\mathbf{h}^H_k\mathbf{w}_j\rvert^2 + \sigma_n^2}}
	\end{flalign}
	and
	\begin{flalign}	
		\begin{aligned}
			\frac{\partial r_k}{\partial \mathbf{w}^*_j} = \frac{2}{1+\gamma_k}\cdot\frac{-\lvert\mathbf{h}^H_k\mathbf{w}_k\rvert^2\mathbf{h}_k^H\mathbf{w}_j\mathbf{h}_k}{\left({\sum_{j\neq k}\lvert\mathbf{h}^H_k\mathbf{w}_j\rvert^2 + \sigma_n^2}\right)^2}, \forall j \neq k.
		\end{aligned}		
	\end{flalign}
	Then, the partial derivative with respect to the precoding matrix $\mathbf{W}$ is given as
	\begin{flalign}	
		\frac{\partial r_k}{\partial \mathbf{W}^*} = \left[\frac{\partial r_k}{\partial \mathbf{w}_1^*},\dots,\frac{\partial r_k}{\partial \mathbf{w}_k^*}\right].
	\end{flalign}
	Besides, for the partial derivative with respect to $\boldsymbol{\phi}$, we first obtain the Jacobian matrix
	\begin{flalign}	
		\mathcal{D}_{\boldsymbol{\phi}^*}\mathbf{h}_k = \mathbf{G}^T\operatorname{diag}(\mathbf{h}^H_{r,k}) 
	\end{flalign}
	and the partial derivative
	\begin{flalign}	
		\frac{\partial \gamma_k}{\partial \mathbf{h}_k}\!=\! \frac{\mathbf{w}_k^H\mathbf{h}_k\mathbf{w}_k}{{\sum_{j\neq k}\lvert\mathbf{h}^H_k\mathbf{w}_j\rvert^2 + \sigma_n^2}}\!-\! \sum_{j\neq k}\frac{\lvert\mathbf{h}^H_k\mathbf{w}_k\rvert^2\mathbf{w}_j^H\mathbf{h}_k\mathbf{w}_j}{\left({\sum_{j\neq k}\lvert\mathbf{h}^H_k\mathbf{w}_j\rvert^2 + \sigma_n^2}\right)^2}.
	\end{flalign}
	Exploiting the transformation $\mathcal{D}_{\mathbf{Z}^*}f(\mathbf{Z}) = \operatorname{vec}^T\left(\frac{\partial}{\partial \mathbf{Z}^*}f(\mathbf{Z})\right)$, we can obtain the partial derivative as
	\begin{flalign}	
		\frac{\partial r_k}{\partial \boldsymbol{\phi}^*}=\frac{\partial r_k}{\partial \gamma_k}\cdot\frac{\partial \gamma_k}{\partial \boldsymbol{\phi}^*} = \frac{2}{1+\gamma_k}\cdot\frac{\partial \gamma_k}{\partial \boldsymbol{\phi}^*},
	\end{flalign}
	where
	\begin{flalign}	
		\frac{\partial \gamma_k}{\partial \boldsymbol{\phi}^*}=\left(\mathcal{D}_{\boldsymbol{\phi}^*}\mathbf{h}_k\right)^T\cdot\frac{\partial \gamma_k}{\partial \mathbf{h}_k}.
	\end{flalign}
	
	For calculating the Euclidean gradients with respect to the MAPs, we can first obtain the partial derivatives with respect to the channels $\mathbf{G}$ and $\mathbf{f}_k,\forall k \in \mathcal{K}$ as
	\begin{align}	
		\frac{\partial \gamma_k}{\partial \mathbf{G}^*}=& \frac{\mathbf{h}_k^H\mathbf{w}_k\operatorname{diag}(\mathbf{f}^H_k)\boldsymbol{\phi}\mathbf{w}_k^H}{{\sum_{j\neq k}\lvert\mathbf{h}^H_k\mathbf{w}_j\rvert^2 + \sigma_n^2}}-\nonumber\\ &\qquad\qquad\sum_{j\neq k}\frac{\lvert\mathbf{h}_k^H\mathbf{w}_k\rvert^2\mathbf{h}_k^H\mathbf{w}_j\operatorname{diag}(\mathbf{f}^H_k)\boldsymbol{\phi}\mathbf{w}_j^H}{\left({\sum_{j\neq k}\lvert\mathbf{h}^H_k\mathbf{w}_j\rvert^2 + \sigma_n^2}\right)^2}
	\end{align}
	and
	\begin{align}	
		\frac{\partial \gamma_k}{\partial \mathbf{f}_k^*}=& \frac{\mathbf{w}_k^H\mathbf{h}_k\operatorname{diag}(\boldsymbol{\phi}^H)\mathbf{G}\mathbf{w}_k^H}{{\sum_{j\neq k}\lvert\mathbf{h}^H_k\mathbf{w}_j\rvert^2 + \sigma_n^2}}-\nonumber\\ &\qquad\qquad\sum_{j\neq k}\frac{\lvert\mathbf{h}_k^H\mathbf{w}_k\rvert^2\mathbf{w}^H_j\mathbf{h}_k\operatorname{diag}(\boldsymbol{\phi}^H)\mathbf{G}\mathbf{w}_j^H}{\left({\sum_{j\neq k}\lvert\mathbf{h}^H_k\mathbf{w}_j\rvert^2 + \sigma_n^2}\right)^2}
	\end{align}
	Note that $\mathbf{t}$ is variable of $\mathbf{G}$, while $\mathbf{u}$ is variable of both $\mathbf{G}$ and $\mathbf{f}_k$. For $\mathbf{t}$, we can first obtain the partial derivatives
	\begin{flalign}	
		\label{T1}
		\frac{\partial \mathbf{G}^*[n,m]}{\partial {t}_m}=\sum_{l=1}^{L} -\jmath\frac{2\pi}{\lambda}\sigma_{G,l}^*e^{\jmath\frac{2\pi}{\lambda}(\boldsymbol{\rho}_{r,l}^T\mathbf{u}_n-{\rho}_{t,l}{t}_m)}{\rho}_{t,l}.
	\end{flalign}
	Take it as the $m$-th element, the partial derivative of $\mathbf{G}^*[n,m]$ with respect to $\mathbf{t}$ can be obtained as
	\begin{equation}
		\frac{\partial \mathbf{G}^*[n,m]}{\partial \mathbf{t}}=\left[0,\dots,0,\frac{\partial \mathbf{G}^*[n,m]}{\partial t_m},0,\dots,0\right]^T.
	\end{equation}
	Then, the $(N(m-1)+n)$-th row of Jacobian matrix $\mathcal{D}_{\mathbf{t}}\mathbf{G}^*$ can be obtained as $({\partial \mathbf{G}^*[n,m]}/{\partial \mathbf{t}})^T$. Finally, the partial derivative of rate with respect to $\mathbf{t}$ is
	\begin{flalign}		
		\label{T3}	
		\operatorname{vec}\left(\frac{\partial r_k}{\partial \mathbf{t}}\right)=\mathcal{D}_{\mathbf{t}}r_k= \frac{2}{1+\gamma_k}\operatorname{vec}^T\left(\frac{\partial \gamma_k}{\partial \mathbf{G}^*}\right)\mathcal{D}_{\mathbf{T}}\mathbf{G}^*.
	\end{flalign}
	
	For $\mathbf{u}$, we can calculate the partial derivatives
	\begin{flalign}	
		\frac{\partial \mathbf{G}^*[n,m]}{\partial \mathbf{u}_n}=\sum_{l=1}^{L} \jmath\frac{2\pi}{\lambda}\sigma_{G,l}^*e^{\jmath\frac{2\pi}{\lambda}(\boldsymbol{\rho}_{r,l}^T\mathbf{u}_n-{\rho}_{t,l}{t}_m)}\boldsymbol{\rho}_{r,l},
	\end{flalign}
	and
	\begin{flalign}	
		\frac{\partial \mathbf{f}_k^*[n]}{\partial \mathbf{u}_n}=-\sum_{l=1}^{L} \jmath\frac{2\pi}{\lambda}\sigma_{k,l}^*e^{\jmath\frac{2\pi}{\lambda}(-\boldsymbol{\rho}_{r,l}^T\mathbf{u}_n)}\boldsymbol{\rho}_{r,l},
	\end{flalign}
	then we can obtain the vectorization of the partial derivative as
	\begin{flalign}	
		\operatorname{vec}\left(\frac{\partial r_k}{\partial \mathbf{u}}\right)=\mathcal{D}_{\mathbf{G}^*}r_k\mathcal{D}_{\mathbf{u}}\mathbf{G}^*+\mathcal{D}_{\mathbf{f}_k^*}r_k\mathcal{D}_{\mathbf{u}}\mathbf{f}_k^*,
	\end{flalign}
	where each term can be calculated similarly as \eqref{T1}-\eqref{T3}.
	
	For the inequality constraints $h_i(\mathbf{X}),\forall i \in \mathcal{D}_I$, we can calculate the partial derivatives 
	\begin{equation}
		\frac{\partial h_i(\mathbf{X})}{\partial \mathbf{u}_{n}} = \bigl((\mathbf{u}_{n}-\mathbf{u}_{n'})^T(\mathbf{u}_{n}-\mathbf{u}_{n'})\bigr)^{-1/2}(\mathbf{u}_{n}-\mathbf{u}_{n'})
	\end{equation}
	and
	\begin{equation}
		\frac{\partial h_i(\mathbf{X})}{\partial \mathbf{u}_{n'}} =- \bigl((\mathbf{u}_{n}-\mathbf{u}_{n'})^T(\mathbf{u}_{n}-\mathbf{u}_{n'})\bigr)^{-1/2}(\mathbf{u}_{n}-\mathbf{u}_{n'}),
	\end{equation}
	which are the $n$-th and $n'$-th 2 elements of $\partial h_{n,n'}(\mathbf{u})/\partial \mathbf{u}$, respectively, while other columns of which are $\mathbf{0}$. For the inequality constraints $h_i(\mathbf{X}),\forall i \in \mathcal{D}_B$, the $m$-th and the $m'$-th elements of the partial derivatives ${\partial h_i(\mathbf{X})}/{\partial \mathbf{t}}$ can be obtained as
	\begin{equation}
		\frac{\partial h_i(\mathbf{X})}{\partial t_{m}} = -\operatorname{sign}(t_m-t_{m'})
	\end{equation}
	and
	\begin{equation}
		 \frac{\partial h_i(\mathbf{X})}{\partial t_{m'}} = \operatorname{sign}(t_m-t_{m'}),
	\end{equation}
	where $\operatorname{sign}(\cdot)$ is the signum function, while other elements of ${\partial h_i(\mathbf{X})}/{\partial \mathbf{t}}$ are $0$.
	
	\section{Convergence Analysis of \textbf{Algorithm 3}}
	If the sequence $\{\mathbf{X}^l\}$ produced by \textbf{Algorithm 2} admits a feasible limit point $\bar{\mathbf{X}}$ satisfying the LICQ conditions, there exists $l_1$ such that for any $l>l_1$, $i\in\mathcal{I}\setminus \mathcal{A}(\bar{\mathbf{X}})$, $h_i(\mathbf{Q})< c$ for some constant $c<0$. Denote the Riemannian gradient of the $l$-th iteration as 
	\begin{flalign}	
		\operatorname{grad}_{\mathbf{X}}g(\mathbf{X}^l)\!=\!\operatorname{grad}_{\mathbf{X}}f(\mathbf{X}^l) + \rho^l \sum_{i\in\mathcal{I}}\lambda_i \operatorname{grad}_{\mathbf{X}}h_i(\mathbf{X}^l)
	\end{flalign}
	where
	\begin{flalign}	
		\lambda_i^l = \frac{e^{h_i(\mathbf{X}^{l})/u^l}}{1+e^{h_i(\mathbf{X}^{l})/u^l}}.
	\end{flalign}
	We can obtain the limit point $\bar{\lambda}_i =0,\forall i \in \mathcal{I}$ since $\lambda_i^l \rightarrow 0 $ as $u^l \rightarrow u_{min}=0$. With sufficient large $l$ and fixed feasible $\bar{\rho}$, we define
	\begin{flalign}	
		\boldsymbol{\nu} = \operatorname{grad}_{\mathbf{X}} f(\bar{\mathbf{X}}) + \bar{\rho} \sum_{i\in\mathcal{I}\cap\mathcal{A}(\bar{\mathbf{X}})}\bar{\lambda}_i \operatorname{grad}h_i(\bar{\mathbf{X}}),
	\end{flalign}
	then the following comparison exists:
	\begin{align}	
		\label{c1}
		\lVert&\boldsymbol{\nu}\rVert \leq \biggl\|\operatorname{grad}_{\mathbf{X}} f(\bar{\mathbf{Q}}) - \mathcal{T}_{\bar{\mathbf{X}}}\operatorname{grad}_{\mathbf{X}} f(\mathbf{X}^l)\biggr\| + \nonumber\\& \biggl\|\bar{\rho} \sum_{i\in\mathcal{N}\cap\mathcal{A}(\bar{\mathbf{X}})}\bar{\lambda}_i \left(\operatorname{grad}_{\mathbf{X}}h_i(\bar{\mathbf{X}})-\mathcal{T}_{\bar{\mathbf{X}}}\operatorname{grad}_{\mathbf{X}}h_i(\mathbf{X}^l)\right)\biggr\|+\nonumber\\&\biggl\|\mathcal{T}_{\bar{\mathbf{X}}}\operatorname{grad}_{\mathbf{X}}f(\mathbf{X}^l) + \bar{\rho}\!\!\!\!\sum_{i\in\mathcal{N}\cap\mathcal{A}(\bar{\mathbf{X}})}\!\!\!\!\bar{\lambda}_i\mathcal{T}_{\bar{\mathbf{X}}}\operatorname{grad}_{\mathbf{X}}h_i(\mathbf{X}^l)\biggr\|.
	\end{align}
	
	If $\mathbf{X}^l \rightarrow \bar{\mathbf{X}}$ with $l\rightarrow \infty$, as given in \cite{rep}, we have
	\begin{equation}
		\label{c2}
		\lim_{l\rightarrow\infty} \left\|\mathcal{T}_{\bar{\mathbf{X}}}\operatorname{grad}_{\mathbf{X}}g({\mathbf{X}^l})-\operatorname{grad}_{\mathbf{X}}g(\bar{\mathbf{X}})\right\|=0,
	\end{equation}
	then the first term of \eqref{c1} tends to zero as $l\rightarrow\infty$. Exploiting the isometry of transportation and linearity \cite{iso}, the second term can be handled as
	\begin{align}	
		\label{c3}
		&\biggl\|\mathcal{T}_{\bar{\mathbf{X}}}\operatorname{grad}_{\mathbf{X}}f(\mathbf{X}^l) + \bar{\rho}\!\!\!\sum_{i\in\mathcal{I}\cap\mathcal{A}(\bar{\mathbf{X}})}\!\!\!\bar{\lambda}_i\mathcal{T}_{\bar{\mathbf{X}}}\operatorname{grad}_{\mathbf{X}}h_i(\mathbf{X}^l)\biggr\|\nonumber\\=&\biggl\|\operatorname{grad}_{\mathbf{X}}f(\mathbf{X}^l) + \bar{\rho}\sum_{i\in\mathcal{I}\cap\mathcal{A}(\mathbf{X}^l)}\bar{\lambda}_i\operatorname{grad}_{\mathbf{X}}h_i(\mathbf{X}^l)\biggr\|\nonumber\\&\leq \biggl\|\operatorname{grad}_{\mathbf{X}}f(\mathbf{X}^l) + \bar{\rho}\sum_{i\in\mathcal{I}}\lambda^l_i\operatorname{grad}_{\mathbf{X}}h_i(\mathbf{X}^l)\biggr\|\nonumber\\&+\biggl\| \bar{\rho}\sum_{i\in\mathcal{I}\cap\mathcal{A}(\mathbf{X}^l)}\left(\bar{\lambda}_i-\lambda_i^l\right)\operatorname{grad}_\mathbf{X}h_i(\mathbf{X}^l)\biggr\|\nonumber\\&+\biggl\| \bar{\rho}\sum_{i\in\mathcal{I}\setminus\mathcal{A}(\mathbf{X}^l)}\lambda_i^l\operatorname{grad}_{\mathbf{X}}h_i(\mathbf{X}^l)\biggr\|.
	\end{align}
	Observing the right hand side of the inequality \eqref{c3}, as $l\rightarrow\infty$, the first term tends to $d_{min}=0$, the second term tends to zero since $\lambda_i^l\rightarrow\bar{\lambda}_i$ as $l\rightarrow\infty$, and the third term tends to zero since $\lambda_i^l\rightarrow0$. Combining \eqref{c3}, \eqref{c2}, and \eqref{c1}, we can conclude that $\lVert\boldsymbol{\nu}\rVert = 0$. Then we can deduce that Lagrange multipliers $\lambda^*_i$ exist, leading to
	\begin{equation}
		\begin{cases}
			\bigl\|\operatorname{grad}_\mathbf{X} f(\bar{\mathbf{X}}) + \sum_{i\in\mathcal{I}}\lambda^*_i \operatorname{grad}_{\mathbf{X}}h_i(\bar{\mathbf{X}})\bigr\| =0 ,\\
			h_i(\bar{\mathbf{X}}) \leq 0, \forall i \in \mathcal{I}\\
			\lambda^*_i \geq 0, \forall i \in \mathcal{I},\\
			\lambda^*_ih_i(\bar{\mathbf{X}}) = 0, \forall i \in \mathcal{I}.
		\end{cases}
	\end{equation}
	Thus, $\bar{\mathbf{X}}$ satisfies the KKT conditions for \eqref{Qsr3}, the proof completes.
	
	\bibliographystyle{IEEEtran}
	\bibliography{Refs}{}

\begin{thebibliography}{10}
\providecommand{\url}[1]{#1}
\csname url@samestyle\endcsname
\providecommand{\newblock}{\relax}
\providecommand{\bibinfo}[2]{#2}
\providecommand{\BIBentrySTDinterwordspacing}{\spaceskip=0pt\relax}
\providecommand{\BIBentryALTinterwordstretchfactor}{4}
\providecommand{\BIBentryALTinterwordspacing}{\spaceskip=\fontdimen2\font plus
\BIBentryALTinterwordstretchfactor\fontdimen3\font minus
  \fontdimen4\font\relax}
\providecommand{\BIBforeignlanguage}[2]{{%
\expandafter\ifx\csname l@#1\endcsname\relax
\typeout{** WARNING: IEEEtran.bst: No hyphenation pattern has been}%
\typeout{** loaded for the language `#1'. Using the pattern for}%
\typeout{** the default language instead.}%
\else
\language=\csname l@#1\endcsname
\fi
#2}}
\providecommand{\BIBdecl}{\relax}
\BIBdecl

\bibitem{6g}
C.-X. Wang \emph{et~al.}, ``{On the road to 6G: Visions, requirements, key
  technologies, and testbeds},'' \emph{IEEE Commun. Surv. Tutor.}, vol.~25,
  no.~2, pp. 905--974, 2023.

\bibitem{mimo1}
Z.~Wang \emph{et~al.}, ``{Extremely large-scale MIMO: Fundamentals, challenges,
  solutions, and future directions},'' \emph{IEEE Wirel. Commun.}, vol.~31,
  no.~3, pp. 117--124, 2024.

\bibitem{mimo2}
S.~A. Busari \emph{et~al.}, ``{Millimeter-wave massive MIMO communication for
  future wireless systems: A survey},'' \emph{IEEE Commun. Surv. Tutor.},
  vol.~20, no.~2, pp. 836--869, 2017.

\bibitem{IRS1}
Q.~Wu and R.~Zhang, ``{Towards smart and reconfigurable environment:
  Intelligent reflecting surface aided wireless network},'' \emph{IEEE Commun.
  Mag.}, vol.~58, no.~1, pp. 106--112, 2020.

\bibitem{IRS2}
Z.~Wan, Z.~Gao, F.~Gao, M.~D. Renzo, and M.-S. Alouini, ``{Terahertz massive
  MIMO with holographic reconfigurable intelligent surfaces},'' \emph{IEEE
  Trans. Commun.}, vol.~69, no.~7, pp. 4732--4750, 2021.

\bibitem{IRS3}
Q.~Wu \emph{et~al.}, ``{Intelligent surfaces empowered wireless network: Recent
  advances and the road to 6G},'' \emph{arXiv preprint arXiv:2312.16918}, 2023.

\bibitem{IRS4}
H.~Guo, Y.-C. Liang, J.~Chen, and E.~G. Larsson, ``{Weighted sum-rate
  maximization for reconfigurable intelligent surface aided wireless
  networks},'' \emph{IEEE Trans. Wirel. Commun.}, vol.~19, no.~5, pp.
  3064--3076, 2020.

\bibitem{IRS5}
Q.~Wu and R.~Zhang, ``{Intelligent reflecting surface enhanced wireless network
  via joint active and passive beamforming},'' \emph{IEEE Trans. Wirel.
  Commun.}, vol.~18, no.~11, pp. 5394--5409, 2019.

\bibitem{IRS6}
J.~Qiu, J.~Yu, A.~Dong, and K.~Yu, ``{Joint beamforming for IRS-aided
  multi-cell MISO system: Sum rate maximization and SINR balancing},''
  \emph{IEEE Trans. Wirel. Commun.}, vol.~21, no.~9, pp. 7536--7549, 2022.

\bibitem{IRS7}
D.~Li, ``{Fairness-aware multiuser scheduling for finite-resolution intelligent
  reflecting surface-assisted communication},'' \emph{IEEE Commun. Lett.},
  vol.~25, no.~7, pp. 2395--2399, 2021.

\bibitem{AIRS}
H.~Lu, Y.~Zeng, S.~Jin, and R.~Zhang, ``Aerial intelligent reflecting surface:
  Joint placement and passive beamforming design with 3d beam flattening,''
  \emph{IEEE Trans. Wirel. Commun.}, vol.~20, no.~7, pp. 4128--4143, 2021.

\bibitem{MA1}
L.~Zhu, W.~Ma, and R.~Zhang, ``{Movable antennas for wireless communication:
  Opportunities and challenges},'' \emph{IEEE Commun. Mag.}, vol.~62, no.~6,
  pp. 114--120, 2024.

\bibitem{MA2}
J.~Zheng \emph{et~al.}, ``{Flexible-position MIMO for wireless communications:
  Fundamentals, challenges, and future directions},'' \emph{IEEE Wirel.
  Commun.}, vol.~31, no.~5, pp. 18--26, 2024.

\bibitem{mimo3}
E.~G. Larsson, O.~Edfors, F.~Tufvesson, and T.~L. Marzetta, ``{Massive MIMO for
  next generation wireless systems},'' \emph{IEEE Commun. Mag.}, vol.~52,
  no.~2, pp. 186--195, 2014.

\bibitem{mimo4}
M.~Z. Chowdhury, M.~Shahjalal, S.~Ahmed, and Y.~M. Jang, ``{6G wireless
  communication systems: Applications, requirements, technologies, challenges,
  and research directions},'' \emph{IEEE Open J. Commun. Soc.}, vol.~1, pp.
  957--975, 2020.

\bibitem{MAIRSISAC}
H.~Wu, H.~Ren, C.~Pan, and Y.~Zhang, ``Movable antenna-enabled ris-aided
  integrated sensing and communication,'' \emph{IEEE Trans. Cogn. Commun.
  Netw.}, Early Access, 2025.

\bibitem{MA3}
L.~Zhu, W.~Ma, and R.~Zhang, ``{Modeling and performance analysis for movable
  antenna enabled wireless communications},'' \emph{IEEE Trans. Wirel.
  Commun.}, vol.~23, no.~6, pp. 6234--6250, 2024.

\bibitem{MA9}
{W. MA, L. Zhu and R. Zhang}, ``{MIMO capacity characterization for movable
  antenna systems},'' \emph{IEEE Trans. Wirel. Commun.}, vol.~23, no.~4, pp.
  3392--3407, 2024.

\bibitem{MA10}
S.~Yang, W.~Lyu, B.~Ning, Z.~Zhang, and C.~Yuen, ``Flexible precoding for
  multi-user movable antenna communications,'' \emph{IEEE Wirel. Commun.
  Lett.}, vol.~13, no.~5, pp. 1404--1408, 2024.

\bibitem{MA11}
H.~Wang, Q.~Wu, and W.~Chen, ``{Movable antenna enabled interference network:
  Joint antenna position and beamforming design},'' \emph{IEEE Wirel. Commun.
  Lett.}, vol.~13, no.~9, pp. 2517--2521, 2024.

\bibitem{MAIRS1}
Y.~Sun, H.~Xu, B.~Ning, Z.~Cheng, C.~Ouyang, and H.~Yang, ``{Sum-rate
  optimization for RIS-aided multiuser communications with movable antenna},''
  \emph{IEEE Wirel. Commun. Lett.}, vol.~14, no.~2, pp. 450--454, 2025.

\bibitem{MAIRS2}
G.~Hu \emph{et~al.}, ``Intelligent reflecting surface-aided wireless
  communication with movable elements,'' \emph{IEEE Wirel. Commun. Lett.},
  vol.~13, no.~4, pp. 1173--1177, 2024.

\bibitem{MAIRS3}
Y.~Zhang, I.~Dey, and N.~Marchetti, ``{RIS-aided wireless communication with
  movable elements geometry impact on performance},'' \emph{arXiv preprint
  arXiv:2405.00141}, 2024.

\bibitem{rtime}
L.~Zhu \emph{et~al.}, ``A tutorial on movable antennas for wireless networks,''
  \emph{IEEE Commun. Surv. Tutor.}, Early Access, 2025.

\bibitem{MA6}
L.~Zhu, W.~Ma, B.~Ning, and R.~Zhang, ``{Movable-antenna enhanced multiuser
  communication via antenna position optimization},'' \emph{IEEE Trans. Wirel.
  Commun.}, vol.~23, no.~7, pp. 7214--7229, 2024.

\bibitem{MA7}
Z.~Cheng, N.~Li, J.~Zhu, and C.~Ouyang, ``{Sum-rate maximization for movable
  antenna enabled multiuser communications},'' \emph{arXiv preprint
  arXiv:2309.11135}, 2023.

\bibitem{MA8}
B.~Feng, Y.~Wu, X.-G. Xia, and C.~Xiao, ``{Weighted sum-rate maximization for
  movable antenna-enhanced wireless networks},'' \emph{IEEE Wirel. Commun.
  Lett.}, vol.~13, no.~6, pp. 1770--1774, 2024.

\bibitem{MAIRS4}
X.~Wei, W.~Mei, Q.~Wu, B.~Ning, and Z.~Chen, ``Movable antennas meet
  intelligent reflecting surface: When do we need movable antennas?'' in
  \emph{Proc. IEEE WCNC}, 2025, pp. 1--6.

\bibitem{MAchan}
W.~Ma, L.~Zhu, and R.~Zhang, ``{Compressed sensing based channel estimation for
  movable antenna communications},'' \emph{IEEE Commun. Lett.}, vol.~27,
  no.~10, pp. 2747--2751, 2023.

\bibitem{CE}
Z.~Xiao \emph{et~al.}, ``Channel estimation for movable antenna communication
  systems: A framework based on compressed sensing,'' \emph{IEEE Trans. Wirel.
  Commun.}, vol.~23, no.~9, pp. 11\,814--11\,830, 2024.

\bibitem{CSM}
Y.~Geng, T.~Hiang~Cheng, K.~Zhong, and K.~Chan~Teh, ``{Unified manifold
  optimization for double-IRS-aided MIMO communication},'' \emph{IEEE Commun.
  Lett.}, vol.~28, no.~7, pp. 1713--1717, 2024.

\bibitem{rmo}
N.~Boumal, \emph{{An Introduction to Optimization on Smooth Manifolds}}.\hskip
  1em plus 0.5em minus 0.4em\relax Cambridge University Press, 2023.

\bibitem{penalty}
J.~Nocedal and S.~J. Wright, \emph{{Numerical Optimization}}.\hskip 1em plus
  0.5em minus 0.4em\relax Springer New York, NY, 2006.

\bibitem{ep}
R.~H. Byrd, J.~Nocedal, and R.~A. Waltz, ``Steering exact penalty methods for
  nonlinear programming,'' \emph{Optimization Methods and Software}, vol.~23,
  no.~2, pp. 197--213, 2008.

\bibitem{rep}
C.~Liu and N.~Boumal, ``{Simple algorithms for optimization on Riemannian
  manifolds with constraints},'' \emph{Appl. Math. Optim.}, vol.~82, no.~3, pp.
  949--981, 2020.

\bibitem{rbfgs}
W.~Huang, P.-A. Absil, and K.~A. Gallivan, ``{A Riemannian BFGS method for
  nonconvex optimization problems},'' in \emph{ENUMATH}.\hskip 1em plus 0.5em
  minus 0.4em\relax Springer, 2016, pp. 627--634.

\bibitem{lbfgs}
P.~T. Boggs and R.~H. Byrd, ``{Adaptive, limited-memory BFGS algorithms for
  unconstrained optimization},'' \emph{SIAM J. Optim.}, vol.~29, no.~2, pp.
  1282--1299, 2019.

\bibitem{lrbfgs}
X.~Yuan, W.~Huang, P.-A. Absil, and K.~A. Gallivan, ``{A Riemannian
  limited-memory BFGS algorithm for computing the matrix geometric mean},''
  \emph{Procedia Computer Science}, vol.~80, pp. 2147--2157, 2016.

\bibitem{conv}
N.~Boumal, P.-A. Absil, and C.~Cartis, ``Global rates of convergence for
  nonconvex optimization on manifolds,'' \emph{IMA J. Numer. Anal.}, vol.~39,
  no.~1, pp. 1--33, 2019.

\bibitem{licq}
W.~H. Yang, L.-H. Zhang, and R.~Song, ``{Optimality conditions for the
  nonlinear programming problems on Riemannian manifolds},'' \emph{Pac. J.
  Optim.}, vol.~10, no.~2, pp. 415--434, 2014.

\bibitem{los}
R.~W. Heath, N.~Gonzalez-Prelcic, S.~Rangan, W.~Roh, and A.~M. Sayeed, ``{An
  overview of signal processing techniques for millimeter wave MIMO systems},''
  \emph{IEEE J. Sel. Top. Signal Process.}, vol.~10, no.~3, pp. 436--453, 2016.

\bibitem{CM}
A.~Hjorungnes and D.~Gesbert, ``{Complex-valued matrix differentiation:
  Techniques and key results},'' \emph{IEEE Trans. Signal Process.}, vol.~55,
  no.~6, pp. 2740--2746, 2007.

\bibitem{iso}
N.~Boumal, \emph{{An Introduction to Optimization on Smooth Manifolds}}.\hskip
  1em plus 0.5em minus 0.4em\relax Cambridge University Press, 2023.

\end{thebibliography}
	
\end{document}